\documentclass[mreferee]{gji}
\usepackage{footnote}
\usepackage{timet}
\usepackage{algorithm}
\usepackage{algorithmic}
\usepackage[pdftex]{graphicx}
\usepackage{epstopdf}
\usepackage{epsfig,subfigure}
\usepackage{epsf}
\usepackage{graphics}
\usepackage{url}
\usepackage{color}
\usepackage{amssymb}
\usepackage{amsmath}

\graphicspath{{./Figures/}}

\addtocounter{footnote}{0}
\newcommand{\refeq[1]}{(\ref{#1})}
\newcommand{\bfm}{\mathbf{m}}
\newcommand{\bfd}{\mathbf{d}}
\newcommand{\bfx}{\mathbf{x}}

\newcommand{\bfr}{\mathbf{r}}
\newcommand{\bfy}{\mathbf{y}}
\newcommand{\bfh}{\mathbf{h}}
\newcommand{\bfz}{\mathbf{z}}
\newcommand{\bfui}{\mathbf{u}_i}
\newcommand{\bfvi}{\mathbf{v}_i}
\newcommand{\bfe}{\mathbf{e}_{t+1}}

\newcommand{\WL}{W_{{L}_1}}
\newcommand{\zetaopt}{\zeta_{\mathrm{opt}}}
\newcommand{\alphaopt}{\alpha_{\mathrm{opt}}}
\newcommand{\bfdo}{\mathbf{d}_{\mathrm{obs}}}

\newcommand{\tbfdo}{\tilde{\mathbf{d}}_{\mathrm{obs}}}
\newcommand{\bfde}{\mathbf{d}_{\mathrm{exact}}}
\newcommand{\bfma}{\mathbf{m}_{\mathrm{apr}}}

\newcommand{\Wd}{W_{\bfd}}
\newcommand{\Gtilde}{\tilde{G}}
\newcommand{\rtilde}{\tilde{ \bfr}}
\newcommand{\Wz}{W_{\mathrm{z}}}

\newcommand{\diag}{\mathrm{diag}}

\newcommand{\Rn}{\mathcal{R}^{n}}
\newcommand{\Rm}{\mathcal{R}^{m}}
\newcommand{\Rmn}{\mathcal{R}^{m \times n}}
\newcommand{\bfeta}{\mbox{\boldmath{$\eta$}}}
\newcommand{\argmin}[1]{\textnormal{arg} \min_{#1}}

\title[$3$-D Projected $L_{1}$ inversion of gravity data ]{3-D Projected $ L_{1}$ inversion of gravity data using truncated unbiased predictive risk estimator for regularization parameter estimation}
\author[S. Vatankhah, R.~A. Renaut, V.~E. Ardestani]{Saeed Vatankhah $^1$, Rosemary A. Renaut $^2$ and Vahid E. Ardestani $^1$ \\
$^1$ Institute of Geophysics, University of Tehran, Iran \\
$^2$ School of Mathematical and Statistical Sciences, Arizona State University, Tempe, AZ, USA.}
\begin{document}
\maketitle

\begin{summary}
Sparse inversion of  gravity data based on $L_1$-norm regularization is discussed. An  iteratively reweighted least squares algorithm is used  to solve the problem.  At each iteration the solution of a linear system of equations and the determination of a suitable regularization parameter are considered. The LSQR iteration is used to project the system of equations onto a smaller subspace that inherits the ill-conditioning of the full space  problem. We show that the gravity kernel is only mildly to moderately ill-conditioned. Thus, while the dominant spectrum of the projected problem accurately approximates the dominant spectrum of the full space problem, the entire spectrum of the projected problem inherits the ill-conditioning of the full problem. Consequently, determining the regularization parameter based on the entire spectrum of the projected problem  necessarily over compensates for the non-dominant portion of the spectrum and leads to inaccurate approximations for the full-space solution. In contrast, finding the regularization parameter using a truncated singular space of the projected operator is efficient and effective. Simulations for synthetic examples with noise demonstrate the approach using the method of unbiased predictive risk estimation for the truncated projected spectrum. The method is used on gravity data from the Mobrun ore body,  northeast of Noranda, Quebec, Canada. The $3$-D reconstructed model is in agreement with known drill-hole information. 
\end{summary}

\section{Introduction}\label{sec:intro}
The gravity data inverse problem is the estimation of the unknown subsurface density and its geometry from a set of  gravity observations measured on the surface. Because the problem is under determined  and non-unique, finding  a stable and geologically plausible solution is feasible only with the imposition of additional information about the model \cite{LiOl:96,PoZh:99}.  Standard methods proceed with the minimization of a global objective function for the model parameters $\bfm$, comprising a data misfit term, $\Phi(\bfm)$,  and stabilizing regularization term, $S(\bfm)$, with balancing provided by a regularization parameter $\alpha$, 
\begin{eqnarray}\label{globalfunction}
P^{\alpha}(\bfm)=\Phi(\bfm)+\alpha^2 S(\bfm). 
\end{eqnarray}
The data misfit measures how well the calculated data reproduces the observed data, typically measured in potential field inversion with respect to a weighted $L_2$-norm  \footnote{\label{pnorm} Throughout we use the standard definition of the $L_{p}$-norm given by $\|\bfx \|_{p}=(\sum_{i=1}^n |x_{i}|^{p})^{\frac{1}{p}}, p\geq 1$, for arbitrary vector $\bfx \in \Rn$, } \cite{LiOl:96,Pi:09}.  Depending on the type of  desired model features to be recovered  through the inversion, there are several choices for the stabilizer, $S(\bfm)$. Imposing $S(\bfm)$ as a $L_2$-norm constraint provides a model with minimal structure.  Depth weighting and low order derivative operators  have been successfully adopted in the geophysical literature \cite[(4)]{LiOl:96}, but the recovered models present with smooth features, especially blurred boundaries, that are not always consistent with  real geological structures \cite{Far:2008}. Alternatively, the minimum volume constraint, \cite{LaKu:83}, and its generalization the minimum support (MS) stabilizer, \cite{PoZh:99,Zhdanov:2002},  yield compact models with sharp interfaces, as do the minimum gradient support (MGS) stabilizer and total variation regularization, which minimize the volume over which the gradient of the model parameters is nonzero, \cite{PoZh:99,Zhdanov:2002,ZhTol,BCO:2002}. Sharp and focused images of the subsurface are also achieved using  $L_1$-norm stabilization \cite{FaOl:98,Far:2008,Loke:2003,SunLi:2014}. With all these constraints \eqref{globalfunction} is  non-linear in $\bfm$ and an iterative algorithm is needed to minimize $P^{\alpha}(\bfm)$. Here we use an iteratively reweighted least squares (IRLS) algorithm in conjunction with  $L_1$-norm stabilization and depth weighting in order to obtain a sparse solution of the gravity inverse problem. 

For small-scale problems, the generalized singular value decomposition (GSVD), or singular value decomposition (SVD),  provide both the  regularization parameter-choice method and the solution minimizing \eqref{globalfunction} in a computationally convenient form \cite{ChNaOl:2008,ChCh:2003}, but are not computationally feasible, whether in terms of computational time or memory, for large scale problems. Alternative approaches to overcome the computational challenge of determining a practical and large scale $\bfm$, include for example  applying  the wavelet transform to compress the sensitivity matrix (Li $\&$ Oldenburg \shortcite{LiOl:03}),   using the symmetry of the gravity forward model to minimize the size of the sensitivity matrix (Boulanger $\&$ Chouteau \shortcite{BoCh:2001}), data-space inversion to yield a system of equations with dimension equal to the number of observations, (Siripunvaraporn $\&$ Egbert  \shortcite{SiEg:00} and Pilkington \shortcite{Pi:09}), and iterative  methods that project  the problem to a smaller subspace (Oldenburg et al. \shortcite{OlMcEl:93}). Our focus here is the use of the iterative LSQR algorithm based on the Lanczos Golub-Kahan bidiagonalization in which a  small Krylov subspace for the solution is generated (Paige $\&$ Saunders \shortcite{PaSa:1982a,PaSa:1982b}). It is analytically equivalent to  applying the conjugate gradient (CG) algorithm but has more favorable analytic properties particularly for ill-conditioned systems, Paige $\&$ Saunders \shortcite{PaSa:1982a}, and has been widely adopted for the solution of regularized least squares problems, e.g. \cite{BJ:96,Hansen:98,Hansen:2007}. Further, using the SVD for the projected system provides  both the solution and the regularization parameter-choice methods in the same computationally efficient form as used for small scale problems and requires little effort beyond the development of the Krylov subspace, \cite{OlLi:94}. 
  
Widely-used approaches for estimating  $\alpha$ include the L-curve \cite{Hansen:92}, Generalized Cross Validation (GCV) \cite{GHW:1979,Mar:1970}, and the Morozov  and $ \chi^2$-discrepancy principles, \cite{Morozov:66} and \cite{MeRe:09,RHM:2010,VRA:2014b}, respectively. Although we have shown in previous investigations of the small scale gravity inverse problem  that the method of unbiased predictive risk estimation  (UPRE), \cite{Vogel:2002}, outperforms  these standard techniques, especially for high noise levels \cite{VAR:2015}, it is not immediate that this conclusion applies when $\alpha$ must be determined for the projected problem.  For example, the GCV method generally overestimates the regularization parameter for the subspace, but introducing a weighted GCV, dependent on a weight parameter, yields  regularization parameters that are more appropriate (Chung et al. \shortcite{ChNaOl:2008}). Our work extends the analysis of the UPRE for the projected problem that was provided in Renaut et al, \shortcite{RVA:2017}. Exploiting the dominant properties of the projected subspace provides an estimate of $\alpha$ using  a truncated application of the UPRE,  denoted by TUPRE, and yields full space solutions that are not under smoothed.

\section{Theory}\label{theory}
\subsection{ Inversion methodology }\label{inversionmethod}
The  $3$-D inversion of gravity data using a linear model is well known, see e.g. \cite{Blakely}.  The subsurface volume is discretized using a set of cubes, in which the cell sizes are kept fixed during the inversion, and the values of densities at the cells  are the model parameters to be determined in the inversion \cite{BoCh:2001,LiOl:98}. For unknown model parameters $\bfm=(\rho_{1}, \rho_{2},\dots, \rho_{n}) \in \Rn$, $\rho_{j}$ the density in cell $j$,  measured data $\bfdo \in \Rm$, and $G$ the sensitivity matrix resulting from the discretization of the forward operator which maps from the model space to the data space,  
 the gravity data satisfy the underdetermined linear system
\begin{eqnarray}\label{d=gm}
\bfdo= G\bfm, \quad G \in \Rmn, \quad m\ll n.
\end{eqnarray} 
The goal of the gravity inverse problem is to find a stable and geologically plausible density model $\bfm$ that reproduces $\bfdo$ at the noise level. We briefly review the stabilized method used here. 

Suppose that $\bfma$ is an initial estimate of the model, possibly
known from a previous investigation, or taken to be zero \cite{LiOl:96}, then the residual and discrepancy from the background data are given by 
\begin{equation}\label{residual}
\bfr=\bfdo-G\bfma \quad \mathrm{and} \quad   \bfy=\bfm-\bfma,
\end{equation}
respectively. Now \refeq[globalfunction] is replaced by  
\begin{eqnarray}\label{globalfunction2}
P^{\alpha}(\bfy)=\| \Wd(G\bfy-\bfr)  \|_2^2 + \alpha^2 \|\bfy \|_1,
\end{eqnarray}
where diagonal matrix $\Wd$ is the data weighting matrix whose $i$th element is the inverse of the standard deviation of the error in the $i$th datum, $\eta_i$, under the assumption that $\bfdo=\bfde+\bfeta$,  and we use an $L_1$-norm  stabilization for $S(\bfm)$.  The $L_1$ norm is approximated using
\begin{eqnarray}\label{WL}
\|\bfy \|_1 \approx \|\WL(\bfy)\bfy\|_2^2, \quad\mathrm{for}\quad 
\left(\WL(\bfy)\right)_{ii}=\frac{1}{(y_{i}^2+\epsilon^2)^{1/4}},\end{eqnarray} 
for very small $\epsilon>0$,  \cite{Voronin:2012,WoRo:07}. As it is also necessary to use a depth weighting matrix to avoid concentration of the model near the surface, see Li $\&$ Oldenburg \shortcite{LiOl:98} and Boulanger $\&$ Chouteau \shortcite{BoCh:2001},  we modify the stabilizer using
\begin{equation}\label{depthmatrix}
W=\WL(\bfy) \Wz, \quad \mathrm{for} \quad \Wz = \diag(z_{j}^{-\beta}),
\end{equation}
where $z_{j}$ is the mean depth of cell $j$ and $\beta$ determines the cell weighting. Then by  \refeq[residual], $\bfm=\bfy+\bfma$, where $\bfy$ minimizes
\begin{eqnarray}\label{globalfunction3}
P^{\alpha}(\bfy)=\| \Wd(G\bfy-\bfr)  \|_2^2 + \alpha^2 \| W\bfy \|_2^2.
\end{eqnarray}
Analytically, assuming  the null spaces of $\Wd G$ and $W$ do not intersect, and $W$ is fixed, the unique minimizer of \refeq[globalfunction3] is given by
\begin{eqnarray}\label{ysolution}
\bfy(\alpha)=(\Gtilde^T\Gtilde+\alpha^2 W^TW)^{-1}\Gtilde^T\rtilde,
\end{eqnarray}
where for ease of presentation we introduce  $\Gtilde=\Wd G$ and $\tilde{\bfr}=\Wd\bfr$. Using the invertibility of diagonal matrix, $W$, (\ref{globalfunction3}) is easily transformed to standard Tikhonov form, see Vatankhah et al. \shortcite{VAR:2015},  
\begin{eqnarray}\label{globalfunctionh}
P^{\alpha}(\bfh)=\| \tilde{\tilde{G}} \bfh- \tilde{\bfr} \|_2^2 + \alpha^2 \|\bfh \|_2^2.
\end{eqnarray}
Here we introduce $\bfh(\alpha)=W\bfy(\alpha)$ and right preconditioning of $\Gtilde$ given by $\tilde{\Gtilde} = \Gtilde W^{-1}$.  
Then, the model update is given by 
\begin{eqnarray}\label{hsolution}
\bfm(\alpha)=\bfma+W^{-1}\bfh(\alpha).
\end{eqnarray}
For small scale problems $\bfh(\alpha) = (\tilde{\tilde{G}}^T\tilde{\tilde{G}}+ \alpha^2 I_n)^{-1} \tilde{\tilde{G}}^T \tilde{\bfr}$ is efficiently obtained using the SVD of  $\tilde{\tilde{G}}$, see Appendix~\ref{svdsolution}. 

Noting now, from \eqref{depthmatrix}, that $\WL$ depends on the model parameters it is immediate that the solution must be obtained iteratively. We use the iteratively reweighted least squares (IRLS) algorithm to obtain the solution \cite{LaKu:83,PoZh:99,Zhdanov:2002,Voronin:2012,WoRo:07}. Matrix  $\WL$ is  updated each iteration using the most recent estimates of the  model parameters, and the IRLS iterations are terminated when either  the solution satisfies the noise level, $\chi_{\mathrm{Computed}}^2 =  \| \Wd(\bfdo -G\bfm^{(k)})\|_2^2 \leq m+\sqrt{2m} $, \cite{BoCh:2001}, or a predefined maximum number of iterations, $K_{\mathrm{max}}$, is reached. At each iterative step any cell density  value that falls outside  practical lower and upper  bounds, $[\rho_{\mathrm{min}},\rho_{\mathrm{max}}]$, is projected back to the nearest constraint value, to assure that reliable subsurface models are recovered. The  IRLS algorithm for small scale  $L_1$ inversion is summarized in Algorithm~\ref{svdalgorithm}.

\begin{figure*}
\subfigure{\label{1a}\includegraphics[width=.4\textwidth, height=0.3\textwidth]{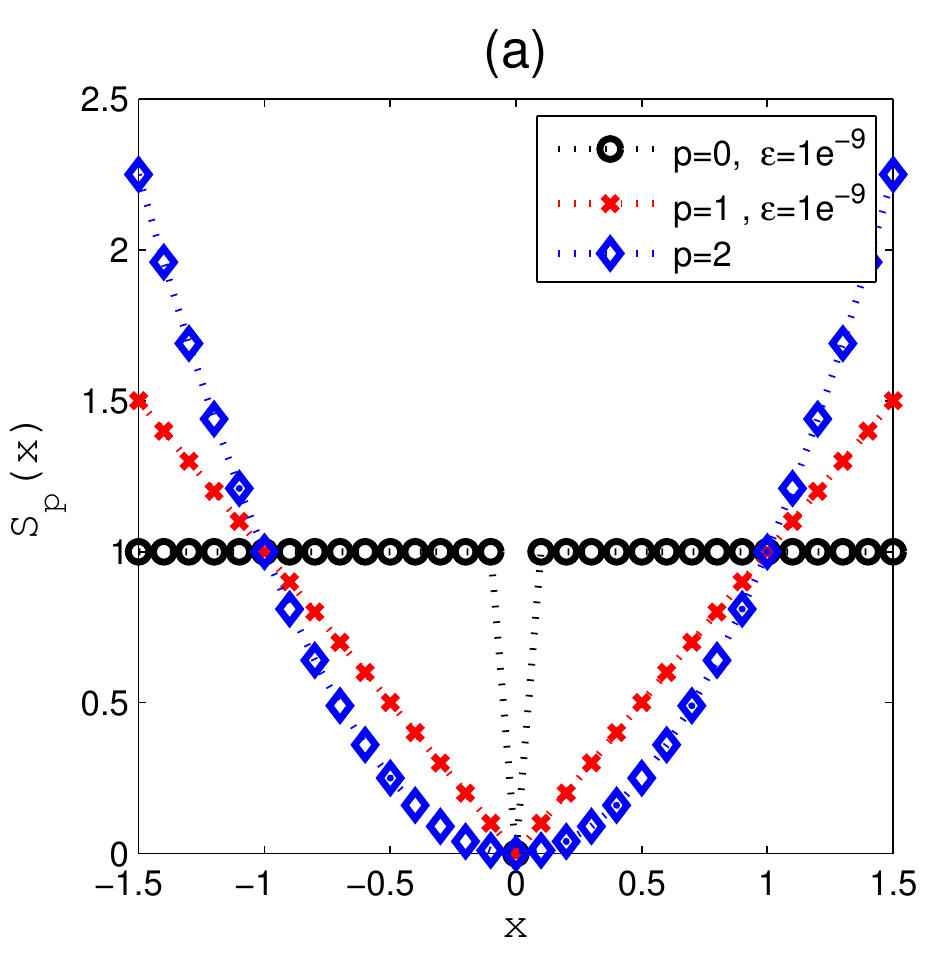}}
\subfigure{\label{1b}\includegraphics[width=.4\textwidth, height=0.3\textwidth]{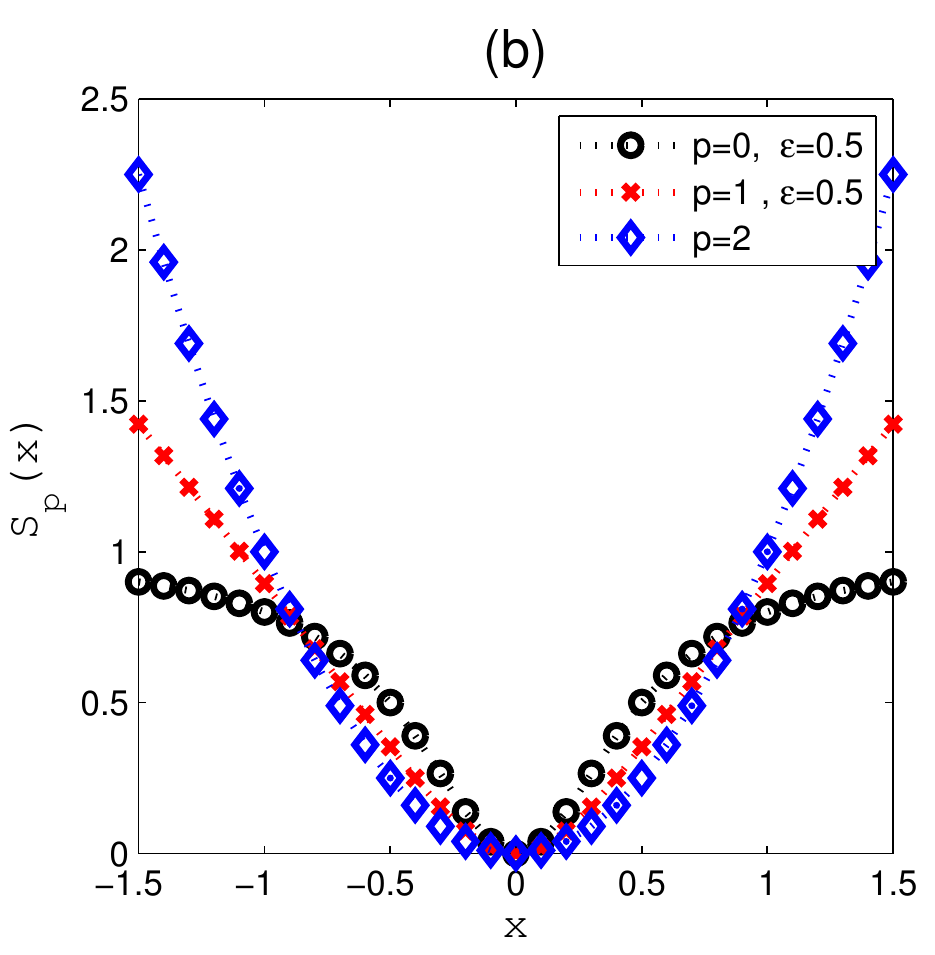}}
\caption {Illustration of different norms for two values of parameter $\epsilon$.  (a) $\epsilon=1e{-9}$; (b) $\epsilon=0.5$.} \label{fig1}
\end{figure*}

In Algorithm~\ref{svdalgorithm} we note that the calculation of $\WL$ depends on a fourth root. To contrast the impact of using different stabilizers we introduce the general formulation 
\begin{eqnarray}\label{generalnorm}
S_p(\bfx)=\sum_{i=1}^n s_p(x_i) 
\quad \mathrm{where} \quad
s_p(x)= \frac{x^2}{{(x^2+\epsilon^2)}^{\frac{2-p}{2}}}.
\end{eqnarray}
When $\epsilon$ is sufficiently small, (\ref{generalnorm}) yields the approximation of  the $L_p$ norm for $p=2$ and $p=1$.  The case with $p=0$,  corresponding to the compactness constraint used in Last $\&$ Kubik \shortcite{LaKu:83}, does not meet the mathematical requirement to be regarded as a norm  and is  commonly used to denote the number of nonzero entries in $\bfx$. Fig.~\ref{fig1} demonstrates the impact of the choice of $\epsilon$ on $s_p(x)$ for $\epsilon = 1e^{-9}$, Fig.~\ref{1a}, and $\epsilon=0.5$, Fig.~\ref{1b}. For larger $p$, more weight is imposed on large  elements of $\bfx$, large elements will be penalized more heavily than small elements during minimization \cite{SunLi:2014}. Hence, as is known,  $L_2$  tends to discourage the occurrence of large elements in the inverted model, yielding smooth models, while  $L_1$ and $L_0$  allow large elements leading to the recovery of blocky features. Note, $s_{0}(x)$ is not quadratic and asymptotes to one away from $0$, regardless of the magnitude of $x$. Hence the penalty on the model parameters does not depend on their relative magnitude, only on whether or not they lie above or below a threshold dependent on $\epsilon$ \cite{Ajo:2007}. While  $L_0$  preserves sparsity better than $L_1$, the solution obtained using $L_0$ is more dependent on the choice of $\epsilon$. The minimum support constraint, $p=0$, can be obtained immediately using  Algorithm~\ref{svdalgorithm}  by replacing the fourth root in the calculation of $\WL$ by the square root.  We return to the estimation of $\alpha^{(k)}$ in Algorithm~\ref{svdalgorithm} step~\ref{start} in Section~\ref{parameter estimation}.

\begin{algorithm}
\caption{Iterative $L_1$ Inversion to find $\bfm(\alpha)$.}\label{svdalgorithm}
\begin{algorithmic}[1]
\REQUIRE $\bfdo$, $\bfma$, $G$, $\Wd$, $\epsilon > 0$, $\rho_{\mathrm{min}}$, $\rho_{\mathrm{max}}$, $K_{\mathrm{max}}$, $\beta$
\STATE Calculate $\Wz$, $\tilde{G}= \Wd G$, and $\tbfdo= \Wd \bfdo$
\STATE Initialize $\bfm^{(0)}=\bfma$,  $W^{(1)}=\Wz$, $k=0$
\STATE Calculate $\tilde{\bfr}^{(1)}=\tbfdo-\tilde{G}\bfm^{(0)}$, $\tilde{\tilde{G}}^{(1)}=\tilde{G}(W^{(1)})^{-1} $ 
\WHILE {$k<K_{\mathrm{max}}$} 
\STATE {$k=k+1$}
\STATE {\label{start}Calculate SVD, $U\Sigma V^T$, of $\tilde{\tilde{G}}^{(k)}$. Find $\alpha^{(k)}$ and update $\bfh^{(k)}=\sum_{i=1}^{m} \frac{\sigma_i^2}{\sigma_i^2+(\alpha^{(k)})^2} \frac{\bfui ^T\tilde{\bfr}^{(k)}}{\sigma_i} \bfvi$.}
\STATE {\label{stop}Set $\bfm^{(k)}=\bfm^{(k-1)}+ (W^{(k)})^{-1}\bfh^{(k)}$}
\STATE {Impose constraint conditions on $\bfm^{(k)}$ to force $\rho_{\mathrm{min}}\le \bfm^{(k)} \le \rho_{\mathrm{max}}$}
\STATE {Test convergence $\| \Wd(\bfdo -G\bfm^{(k)})\|_2^2 \leq m+\sqrt{2m} $.  Exit loop if converged}
\STATE {Calculate the residual $\tilde{\bfr}^{(k+1)}=\tbfdo-\tilde{G}\bfm^{(k)}$}
\STATE {\label{Wupdate}Set $\WL^{(k+1)} =\diag\left( \left((\bfm^{(k)}-\bfm^{(k-1)})^2+\epsilon^2 \right)^{-1/4}\right)$,  as in \refeq[WL], and
$W^{(k+1)}=\WL^{(k+1)}\Wz$}
\STATE {Calculate $\tilde{\tilde{G}}^{(k+1)}=\tilde{G}(W^{(k+1)})^{-1} $}
\ENDWHILE
\ENSURE Solution $\rho=\bfm^{(k)}$. $K=k$.
\end{algorithmic}
\end{algorithm}

\subsection{Application of the LSQR algorithm} \label{GKB}
As already noted, it is not practical to use the SVD for practical large scale problems.  Here we use the Golub Kahan bidiagonalization (GKB) algorithm, see Appendix~\ref{gkbappen}, which is the fundamental step of the LSQR algorithm for solving the damped least squares problem as given in Paige $\&$ Saunders \shortcite{PaSa:1982a,PaSa:1982b}. The solution of the inverse problem is projected to a smaller subspace using $t$ steps of GKB  dependent on the system matrix and the observed data, here $\tilde{\tilde{G}}$ and $\tilde{\bfr}$, respectively. Bidiagonal matrix $ B_t \in \mathcal{R}^{(t+1) \times t}$ and matrices $H_{t+1} \in \mathcal{R}^{m \times (t+1)}$, $A_t \in \mathcal{R}^{n \times t}$ with orthonormal columns are generated such that
\begin{eqnarray}\label{gkb}
\tilde{\tilde{G}}A_t = H_{t+1}B_t, \quad  H_{t+1}\bfe=\tilde{\bfr}/\| \tilde{\bfr}\|_2.
\end{eqnarray}
Here, $\bfe$ is the unit vector of length $t+1$ with a $1$ in the first entry. The columns of $A_{t}$ span the Krylov subspace $\mathcal{K}_{t}$ given by 
\begin{eqnarray}\label{krylov}
\mathcal{K}_t(\tilde{\tilde{G}}^T \tilde{\tilde{G}},\tilde{\tilde{G}}^T\tilde{\bfr})=\mathrm{span} \lbrace \tilde{\tilde{G}}^T\tilde{\bfr},(\tilde{\tilde{G}}^T \tilde{\tilde{G}})\tilde{\tilde{G}}^T\tilde{\bfr}, (\tilde{\tilde{G}}^T \tilde{\tilde{G}})^2\tilde{\tilde{G}}^T\tilde{\bfr}, \dots, (\tilde{\tilde{G}}^T \tilde{\tilde{G}})^{t-1}\tilde{\tilde{G}}^T\tilde{\bfr} \rbrace,
\end{eqnarray}
and an approximate solution $\bfh_t$ that lies in this Krylov subspace will have the form $\bfh_t=A_t \bfz_t$, $\bfz_t \in \mathcal{R} ^{t}$. This Krylov subspace changes for each IRLS iteration $k$. Preconditioner $W$ is not used to accelerate convergence but enforces  regularity on the solution \cite{GaNa:2014}

In terms of the projected space, the global objective function (\ref{globalfunctionh}) is replaced by, see Chung et.al. \shortcite{ChNaOl:2008} and Renaut et al. \shortcite{RVA:2017}, 
\begin{eqnarray}\label{globalfunctionproj}
P^{\zeta}(\bfz)= \| B_t \bfz - \| \tilde{\bfr}\|_2 \bfe \|_2^2 + \zeta^2 \| \bfz \|_2^2.
\end{eqnarray}
Here we use (\ref{gkb}) and the fact that both $A_t$ and $H_{t+1}$ are column orthogonal. Further, the regularization parameter $\zeta$ replaces $\alpha$ to make it explicit that, while  $\zeta$ has the  same role as $\alpha$ as a regularization parameter, we can not assume that the regularization required is the same on the projected and full spaces. Analytically the solution of the projected problem (\ref{globalfunctionproj}) is given by 
\begin{eqnarray}\label{zsolution}
\bfz_t(\zeta)= (B_t^T B_t + \zeta^2 I_t)^{-1} B_t^T \| \tilde{\bfr}\|_2 \bfe. 
\end{eqnarray}
Since the dimensions of $B_t$ are  small as compared to the dimensions of $\tilde{\tilde{G}}$,  $t \ll m$, the solution of the projected problem is obtained efficiently using the SVD, see Appendix~\ref{svdsolution}, and yielding the update  $\bfm_t(\zeta)=\bfma+ W^{-1}  A_t \bfz_t(\zeta)$.

Although $H_{t+1}$ and $A_t$ have orthonormal columns in exact arithmetic, Krylov methods lose orthogonality in finite precision. This means that after a relatively low number of iterations the vectors in $H_{t+1}$ and $A_t$  are no longer orthogonal and the relationship between (\ref{globalfunctionh}) and (\ref{globalfunctionproj}) does not hold. Here we therefore use Modified Gram Schmidt reorthogonalization,  see Hansen \shortcite{Hansen:2007} page $75$,  to maintain the column orthogonality. This is crucial for replicating the dominant spectral properties of $\tilde{\tilde{G}}$ by those of $B_t$. We summarize the steps which are needed for implementation of the projected $L_{1}$ inversion in Algorithm~\ref{projectedalgorithm} for a specified projected subspace size $t$. We emphasize the differences between Algorithms~\ref{svdalgorithm} and \ref{projectedalgorithm}. First the size of the projected space $t$ needs to be given. Then steps \ref{start} to \ref{stop}  in Algorithm~\ref{svdalgorithm} are replaced by steps \ref{projstart} to \ref{projstop} in Algorithm~\ref{projectedalgorithm}.

With respect to memory requirements the largest matrix which needs to be stored is $G$, all other matrices are much smaller and have limited impact on the memory and computational requirements. As already noted in the introduction our focus is not on storage requirements for $G$ but on the LSQR algorithm, thus we note only that diagonal weighting matrices of size $n \times n$ require only $\mathcal{O}(n)$ storage and all actions of multiplication, inversion and transpose are accomplished with component-wise vector operations. With respect to the total cost of the algorithms, it is clear that the costs at a given iteration differ due to the replacement of steps  \ref{start} to \ref{stop} in Algorithm~\ref{svdalgorithm}  by steps \ref{projstart} to \ref{projstop} in Algorithm~\ref{projectedalgorithm}. Roughly the full algorithm requires the SVD for a matrix of size $m \times n$, the generation of the update $ \bfh^{(k)}$, using the SVD, and the estimate of the regularization parameter. Given the SVD, and an assumption that one searches for $\alpha$ over a range of $q$ logarithmically distributed estimates for $\alpha$, as would be expected for a large scale problem the estimate of $\alpha$ is negligible in contrast to the other two steps. For $m \ll n$, the cost is dominated by terms of $\mathcal{O}(n^2m)$ for finding the SVD, \cite[Line 5 of table, p.254] {GoLo:96}. In the projected algorithm the SVD step for $B_t$ and the generation of $\zeta$ are  dominated by terms of $\mathcal{O}(t^3)$. In addition the update $\bfh_t^{(k)}$ is a matrix vector multiply of $\mathcal{O}(mt)$ and generating the factorization is $\mathcal{O}(mnt)$, \cite{PaSa:1982a}. Effectively for $t\ll m$, the dominant term $\mathcal{O}(mnt)$ is actually $\mathcal{O}(mn)$ with $t$ as a scaling, as compared to high cost $\mathcal{O}(n^2m)$. The differences between the costs then increase dependent on the number of iterations $K$ that are required. A more precise estimate of all costs is beyond the scope of this paper, and depends carefully on the implementation used for the SVD and the GKB factorization, both also depending on storage and compression of model matrix $G$.

\begin{algorithm}
\caption{Iterative Projected $L_1$ Inversion Using Golub-Kahan bidiagonalization}\label{projectedalgorithm}
\begin{algorithmic}[1]
\REQUIRE $\bfdo$, $\bfma$, $G$, $\Wd$, $\epsilon > 0$, $\rho_{\mathrm{min}}$, $\rho_{\mathrm{max}}$, $K_{\mathrm{max}}$, $\beta$, $t$.

\STATE Calculate $\Wz$, $\tilde{G}= \Wd G$, and $\tbfdo= \Wd\bfdo$
\STATE Initialize $\bfm^{(0)}=\bfma$,  $\WL^{(1)}=I_n$, $W^{(1)}=\Wz$, $k=0$
\STATE Calculate $\tilde{\bfr}^{(1)}=\tbfdo-\tilde{G}\bfm^{(0)}$, $\tilde{\tilde{G}}^{(1)}=\tilde{G}(W^{(1)})^{-1} $
\WHILE {$k<K_{\mathrm{max}}$} 
\STATE {$k=k+1$}
\STATE {\label{projstart} Calculate factorization: $\tilde{\tilde{G}} ^{(k)} A_t^{(k)} = H_{t+1}^{(k)}B_t^{(k)}$ with $H_{t+1}^{(k)}\bfe=\tilde{\bfr}^{(k)}/\| \tilde{\bfr}^{(k)}\|_2$.  }
\STATE {\label{stepzeta} Calculate SVD, $U\Gamma V^T$, of $B_t^{(k)}$. Find $\zeta^{(k)}$ and update $\bfz_t^{(k)}= \sum_{i=1}^{t} \frac{\gamma_i^2}{\gamma_i^2+(\zeta^{(k)})^2} \frac{\bfui ^T(\| \tilde{\bfr}^{(k)}\|_2\bfe)}{\gamma_i} \bfvi$.}
\STATE {\label{projstop}Set $\bfm^{(k)}=\bfm^{(k-1)}+ (W^{(k)})^{-1}A_t^{(k)}\bfz_t^{(k)}$.}
\STATE {Impose constraint conditions on $\bfm^{(k)}$ to force $\rho_{\mathrm{min}}\le \bfm^{(k)} \le \rho_{\mathrm{max}}$}
\STATE {Test convergence $\| \Wd(\bfdo -G\bfm^{(k)})\|_2^2 \leq m+\sqrt{2m} $.  Exit loop if converged}
\STATE {Calculate the residual $\tilde{\bfr}^{(k+1)}=\tbfdo-\tilde{G}\bfm^{(k)}$}
\STATE {\label{stepWL}Set $\WL^{(k+1)} =\diag\left( \left((\bfm^{(k)}-\bfm^{(k-1)})^2+\epsilon^2 \right)^{-1/4}\right)$,  as in \refeq[WL], and
$W^{(k+1)}=\WL^{(k+1)}\Wz$}
\STATE {Calculate $\tilde{\tilde{G}}^{(k+1)}=\tilde{G}(W^{(k+1)})^{-1} $}
\ENDWHILE
\ENSURE Solution $\rho=\bfm^{(k)}$. $K=k$.
\end{algorithmic}
\end{algorithm}

\subsection{Regularization parameter estimation}\label{parameter estimation}
Algorithms~\ref{svdalgorithm} and \ref{projectedalgorithm} require the determination of a regularization parameter, $\alpha$, $\zeta$, steps \ref{start} and \ref{stepzeta}, respectively. The projected solution  $\bfz_t(\zeta)$  also depends explicitly on the subspace size, $t$. Although we will discuss the effect of choosing different $t$ on the solution, our focus here is not on using existing techniques for finding an optimal subspace size $t_{\mathrm{opt}}$,  see for example the discussions in e.g. \cite{HPS09,RVA:2017}. Instead we wish to find $\zeta$ optimally for a fixed projected problem of size $t$ such that the resulting solution appropriately regularizes the full problem, i.e. so that effectively $\zetaopt \approx \alphaopt$, where $\zetaopt$ and $\alphaopt$ are the optimal regularization parameters for the projected and full problems, respectively. Here, we focus on the method of the UPRE for estimating an optimum regularization parameter in which the derivation of the method for a standard Tikhonov function (\ref{globalfunctionh}) is given in Vogel \shortcite{Vogel:2002} as
\begin{eqnarray}\label{fullupre}
\alpha_{\mathrm{opt}}=\argmin{\alpha}\{U(\alpha):= \|(H(\alpha) -I_m)\tilde{\bfr}\|_2^2 +2\,\mathrm{trace}(H(\alpha)) - m\},
\end{eqnarray}
where $ H(\alpha)=\tilde{\tilde{G}} (\tilde{\tilde{G}}^T\tilde{\tilde{G}}+ \alpha^2 I_n)^{-1} \tilde{\tilde{G}}^T $, and we use that, due to weighting using the inverse square root of  the covariance matrix for the noise, the covariance matrix for the noise in $\tilde{\bfr}$ is $I$. Typically, $\alpha_{\mathrm{opt}}$ is found by evaluating (\ref{fullupre}) for a range of $\alpha$, for example by the SVD see  Appendix~\ref{svdparameter}, with the minimum found within that range of parameter values. For the projected problem, $\zeta_{\mathrm{opt}}$ given by UPRE is obtained as, Renaut et al. \shortcite{RVA:2017},  
\begin{eqnarray}\label{subupre}
\zeta_{\mathrm{opt}}=\argmin{\zeta}\{U(\zeta):= \|(H(\zeta) -I_{t+1}) \| \tilde{\bfr} \|_2 \bfe\|_2^2 +2\,\mathrm{trace}(H(\zeta)) - (t+1)\}, 
\end{eqnarray}
where $ H(\zeta)= B_t (B_t^T B_t + \zeta^2 I_t)^{-1} B_t^T$. As for the full problem, see Appendix~\ref{svdparameter}, the SVD of the matrix $B_t$ can be used to find $\zeta_{\mathrm{opt}}$. 

Now, for small $t$, the singular values of $B_t$ approximate the largest singular values of $\tilde{\tilde{G}}$, however, for larger $t$ the smaller singular values of $B_t$  approximate the  smallest singular values of  $\tilde{\tilde{G}}$, so that there is no immediate one to one alignment between the small singular values of $B_t$ with those of $\tilde{\tilde{G}}$ with increasing $t$. Thus, if the regularized projected problem is to give a good approximation for the regularized full problem,  it is important  that the dominant singular values of $\tilde{\tilde{G}}$ used in estimating $\alphaopt$ are well approximated by those of $B_t$ used in estimating $\zetaopt$. In Section~\ref{synthetic} we show that in some situations \refeq[subupre] does not work well. The modification   that does not use the entire subspace for a given $t$, but rather uses a truncated spectrum from $B_t$ for finding the regularization parameter, assures that the dominant $t_{\mathrm{trunc}}$ terms of the right singular subspace are appropriately regularized.

\section{Synthetic examples}\label{synthetic}
In order to understand the impact of using the LSQR iterative algorithm for solving the large scale inversion problem, it is important to briefly review how the solution of the small-scale problem depends on the noise level in the data and on the parameters of the algorithm, for example $\epsilon$ in the regularization term, the constraints $\rho_{\mathrm{max}}$ and $\rho_{\mathrm{min}}$, and the $\chi^2$ test for convergence.
\subsection{Cube}\label{cube}
\begin{figure*}
\subfigure{\label{2a}\includegraphics[width=.45\textwidth]{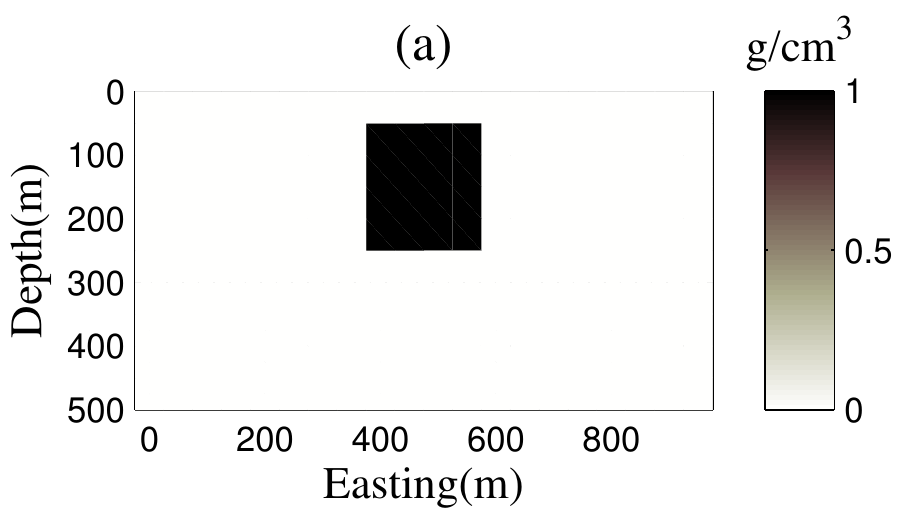}}
\subfigure{\label{2b}\includegraphics[width=.4\textwidth]{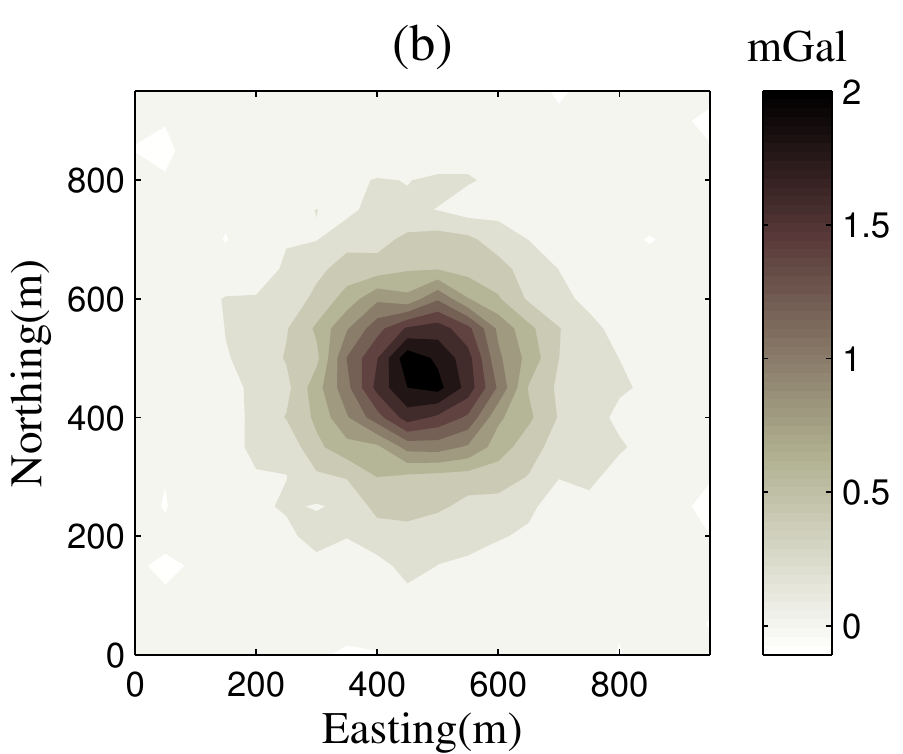}}
\caption { (a) Model of the cube on an homogeneous background. The density contrast of the cube is $1$~g~cm$^{-3}$. (b) Data due to the model and contaminated with  noise $N2$.}
\end{figure*}

We first illustrate the process by which we contrast the $L_1$ algorithms and regularization parameter estimation approaches  for a simple small-scale model that includes a cube with density contrast   $1$~g~cm$^{-3}$ embedded in an homogeneous background.  The cube has size $200$~m in each dimension and is buried at depth $50$~m, Fig.~\ref{2a}. Simulation data on the surface, $\bfde$, are calculated over a $ 20 \times 20 $ regular grid with $50$~m grid spacing. To add noise to the  data, a zero mean Gaussian random matrix $\Theta$ of size $m \times 10$ was generated. Then, setting
\begin{equation}\label{noisydata}
\bfdo^c=\bfde+\left( \tau_1 (\bfde)_i+\tau_2 \|\bfde\|\right) \Theta^c,
\end{equation}
for $c=1:10$, with noise parameter pairs $(\tau_1$, $\tau_2)$, for three choices, $N1: (0.01,0.001)$,  $N2: (0.02,0.005)$ and $N3: (0.03,0.01) $, gives  $10$ noisy right-hand side vectors for each noise level. This noise model is standard in the geophysics literature, see e.g. \cite{LiOl:96}, and incorporates effects of both instrumental and physical noise. We examine the inversion methodology for these different noise levels. We plot the results for one representative right-hand side, at noise level $N2$, Fig.~\ref{2b}, and summarize quantifiable measures of the solutions in tables for all cases at each noise level. 

\label{fig2}
For the inversion  the model region of depth $500$~m is discretized into $ 20 \times 20 \times 10 = 4000$ cells of size $50$~m in each dimension. The background model $\bfma=\mathbf{0}$ and parameters $\beta=0.8$ and $\epsilon^2=1e{-9}$ are chosen for the inversion. Realistic upper and lower density bounds $\rho_{\mathrm{max}}=1$~g~cm$^{-3}$ and $\rho_{\mathrm{min}}=0$~g~cm$^{-3}$, are specified. The iterations are terminated when $\chi_{\mathrm{Computed}}^2 \leq 429 $,  or $k=K_{\mathrm{max}} =50$ is attained. Furthermore,  it was explained by Farquharson \& Oldenburg \shortcite{FaOl:2004} that it is  efficient if the inversion starts with a large value of the regularization parameter. This prohibits imposing excessive structure in the model at early iterations which would otherwise require more iterations to remove artificial structure. In this paper the method introduced by Vatankhah et.al. \shortcite{VAR:2014a,VAR:2015} was used to determine an initial regularization parameter, $\alpha^{(1)}$. Because the non zero singular values $\sigma_i$ of matrix $\tilde{\tilde{G}}$ are known, the initial value 
\begin{equation}\label{initalpha}
\alpha^{(1)}=(n/m)^{3.5}(\sigma_1/\mathrm{mean(\sigma_i)}),
\end{equation}
where the mean is taken over positive $\sigma_i$, can be selected.  For subsequent iterations the UPRE method is used to estimate $\alpha^{(k)}$. 
The results given in the tables are the averages and standard deviations over $10$ samples for  the final iteration $K$, the final regularization parameter $\alpha^{(K)}$ and the relative error of the reconstructed model
\begin{equation}\label{RE}
RE^{(K)}=\frac{\|\bfm_{\mathrm{exact}}-\bfm ^{(K)} \|_2}{\|\bfm_{\mathrm{exact}} \|_2}.
\end{equation}

\subsubsection{Solution using Algorithm~\ref{svdalgorithm}}\label{cubeAlgorithm1}
\begin{figure*}
\subfigure{\label{3a}\includegraphics[width=.45\textwidth]{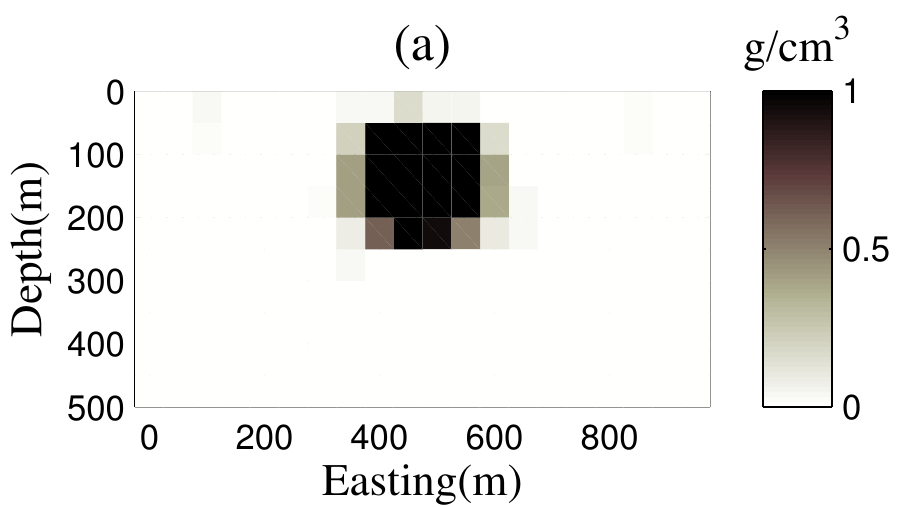}}
\subfigure{\label{3b}\includegraphics[width=.45\textwidth]{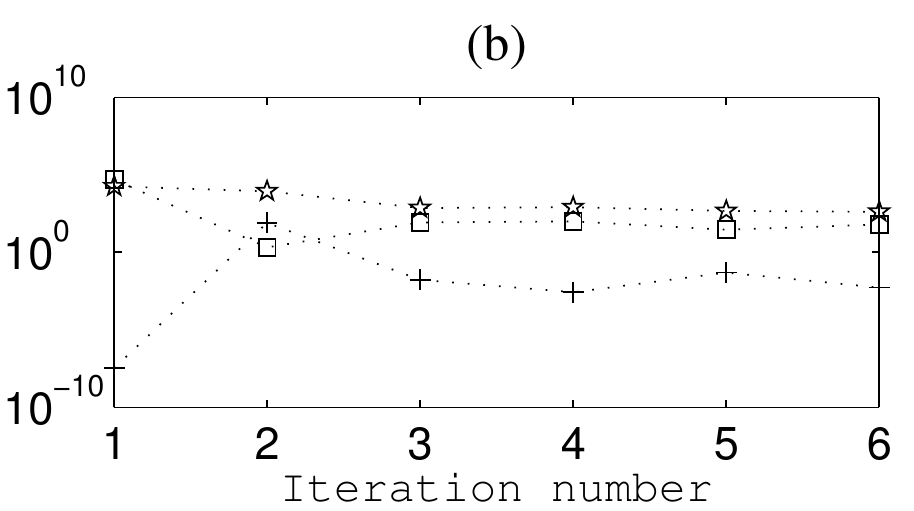}}
\subfigure{\label{3c}\includegraphics[width=.45\textwidth]{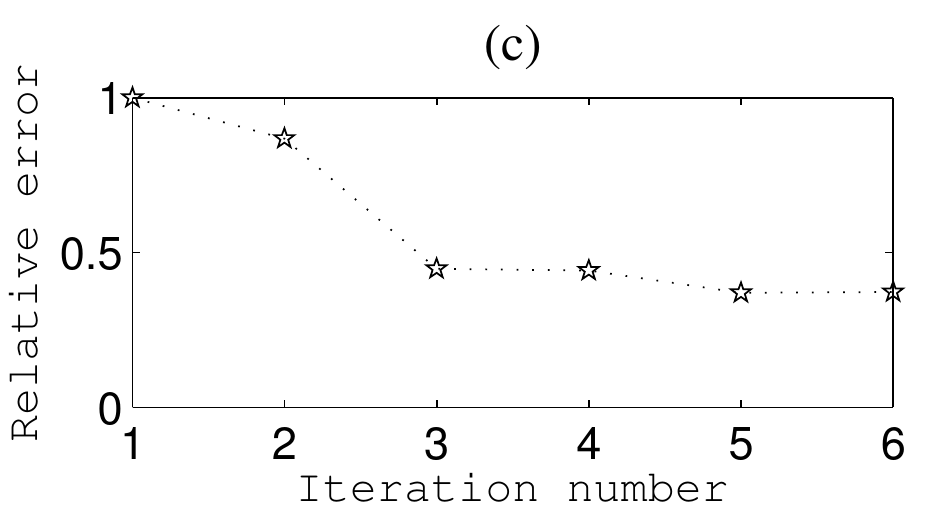}}
\subfigure{\label{3d}\includegraphics[width=.45\textwidth]{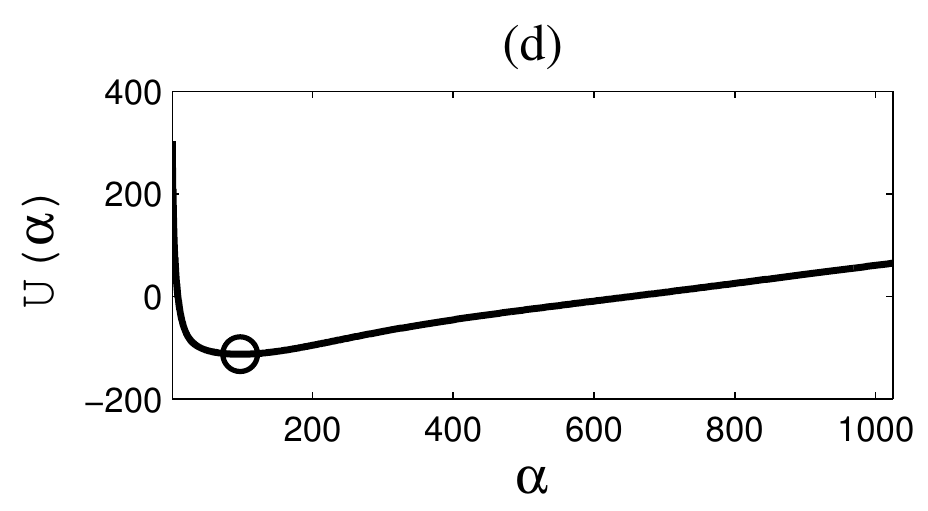}}
\caption {Illustrating use of  Algorithm~\ref{svdalgorithm} with $\epsilon^2=1e{-9}$. UPRE is given for $k=4$. We note that in all examples the figures given are for (a) The reconstructed model;  (b) The progression of the data misfit, $\Phi(\bfm)$ indicated by $\star$, the regularization term,  $ S(\bfm)$  indicated by $+$, and the regularization parameter, $\alpha^{(k)}$  indicated by $\square$, with iteration $k$;  (c) The progression of the relative error $RE^{(k)}$ at each iteration and (d) The UPRE function for a given $k$.} \label{fig3}
\end{figure*}

The results presented in Table~\ref{tab1} are for the $3$ noise levels over $10$ right-hand side data vectors. Convergence of the IRLS is obtained in relatively few iterations, $k<9$, dependent on the noise level, and both $RE$ and $\alpha$ are reasonably robust over the $10$ samples. Results of the inversion for a single sample with noise level $2$  are presented in Fig.~\ref{fig3}, where  Fig.~\ref{3a} shows the reconstructed model, indicating that a focused image of the subsurface is possible using Algorithm~\ref{svdalgorithm}. The constructed models have sharp and distinct interfaces within the embedded medium. The progression of the data misfit $\Phi(\bfm)$, the regularization term $ S(\bfm)$ and regularization parameter $\alpha^{(k)}$ with iteration $k$ are presented in Fig.~\ref{3b}. $\Phi(\bfm)$ is initially large and decays quickly in the first few steps,  but the decay rate decreases dramatically as $k$ increases.  Fig.~\ref{3c} shows the progression of the relative error $RE^{(k)}$ as a function of $k$. There is a dramatic decrease in the relative error for small $k$, after which the error decreases slowly. The UPRE function for iteration $k=4$ is shown in Fig.~\ref{3d}. Clearly, the curves have a nicely defined minimum, which is important in the determination of the regularization parameter. The results presented in the tables are in all cases the average (standard deviation) for $10$ samples of the noise vector. 

\begin{table}
\begin{center}
\caption{The inversion results, for final regularization parameter $\alpha^{(K)}$, relative error $RE^{(K)}$ and number of iterations $K$ obtained by inverting the data from the cube using Algorithm~\ref{svdalgorithm},  with $\epsilon^2=1e{-9}$.}\label{tab1}
\begin{tabular}{c  c  c  c  c }
\hline
Noise&     $\alpha^{(1)}$& $\alpha^{(K)}$& $RE^{(K)}$& $K$  \\ \hline
$N1$ &  47769.1& 117.5(10.6)& 0.318(0.017)& 8.2(0.4)\\
$N2$ &  48623.4& 56.2(8.5)& 0.388(0.023)& 6.1(0.6)\\
$N3$ &  48886.2& 32.6(9.1)& 0.454(0.030)& 5.8(1.3)\\ \hline
\end{tabular}
\end{center}
\end{table} 

\begin{figure*}
\subfigure{\label{4a}\includegraphics[width=.45\textwidth]{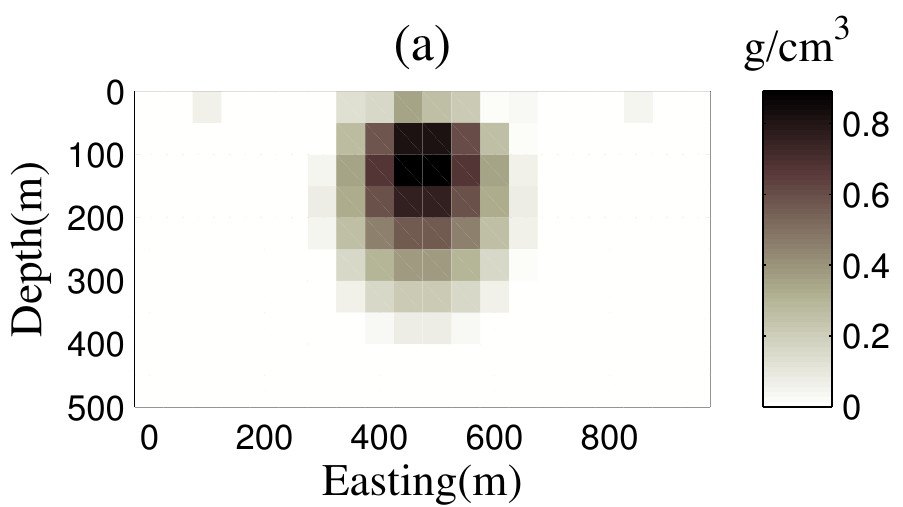}}
\subfigure{\label{4b}\includegraphics[width=.45\textwidth]{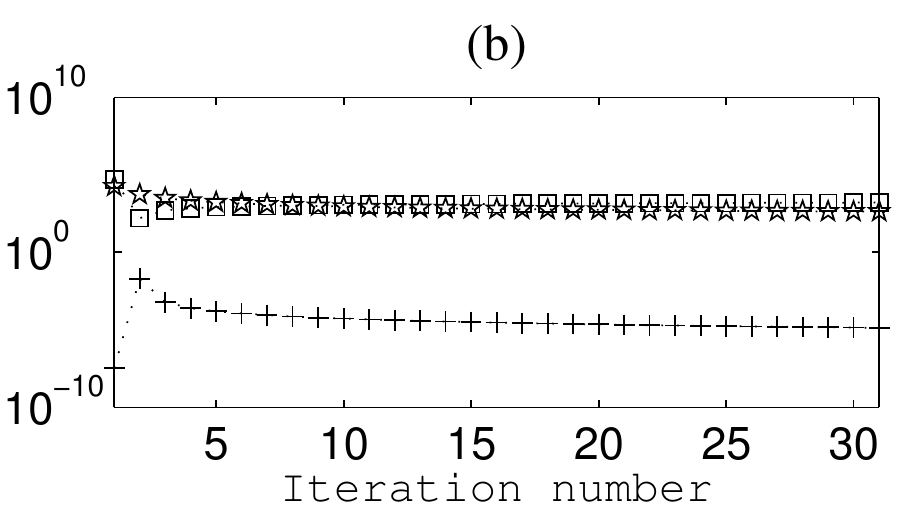}}
\subfigure{\label{4c}\includegraphics[width=.45\textwidth]{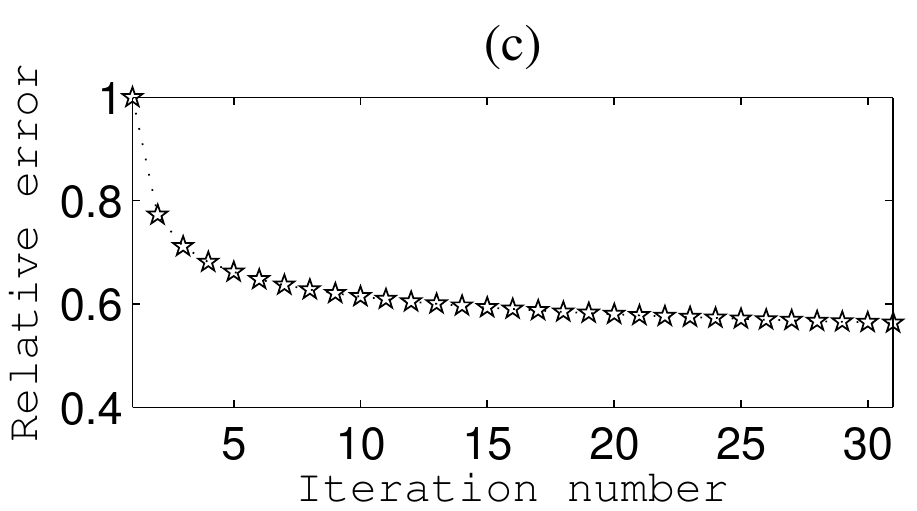}}
\subfigure{\label{4d}\includegraphics[width=.45\textwidth]{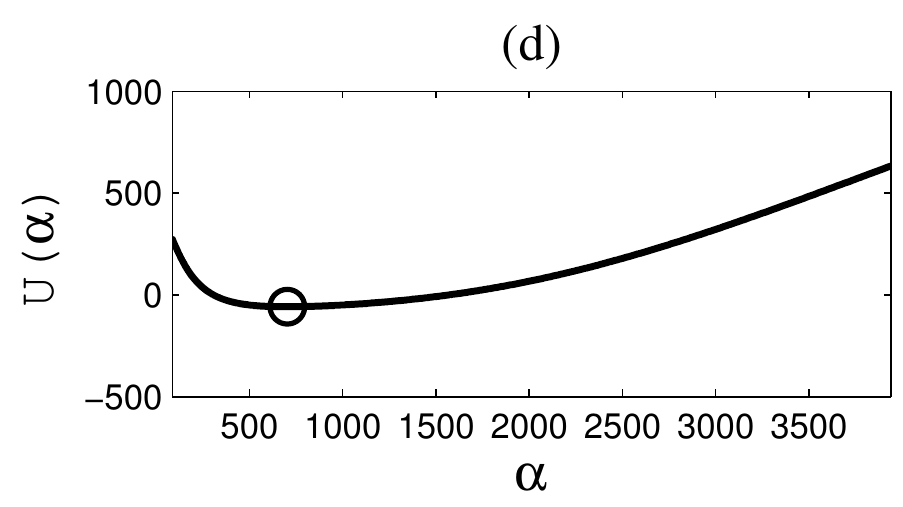}}
\caption {Contrasting use of $\epsilon^2=0.5$ as compared to  $\epsilon^2=1e{-9}$ in Fig.~\ref{fig3}. UPRE is given for $k=4$.} \label{fig4}
\end{figure*}

The role of $\epsilon$ is very important, small values lead to a sparse model that becomes increasingly smooth as $\epsilon$ increases. To determine the dependence  of Algorithm~\ref{svdalgorithm} on other values of $\epsilon^2$, we used $\epsilon ^2=0.5$ and $\epsilon^2=1e{-15}$ with all other parameters chosen  as before. For $\epsilon^2=1e{-15}$ the results, not presented here, are close to those obtained with $\epsilon^2=1e{-9}$. For  $\epsilon ^2=0.5$ the results are significantly different; as presented in Fig.~\ref{fig4} a  smeared-out and fuzzy image of the original model is obtained. The maximum of the obtained density is about $0.85$~g~cm$^{-3}$,  $85\%$ of the imposed $\rho_{\mathrm{max}}$. Note, here, more iterations are needed to terminate the algorithm, $K=31$, and at the final iteration $\alpha^{(31)}=1619.2$ and $RE^{(31)}=0.563$ respectively, larger than their counterparts in the case $\epsilon^2=1e{-9}$.  We found that $\epsilon$ of order $1e{-4}$ to $1e{-8}$ is appropriate for the $L_{1}$ inversion algorithm. Hereafter, we fix $\epsilon^2=1e{-9}$. 

\begin{figure*}  
\subfigure{\label{5a}\includegraphics[width=.45\textwidth]{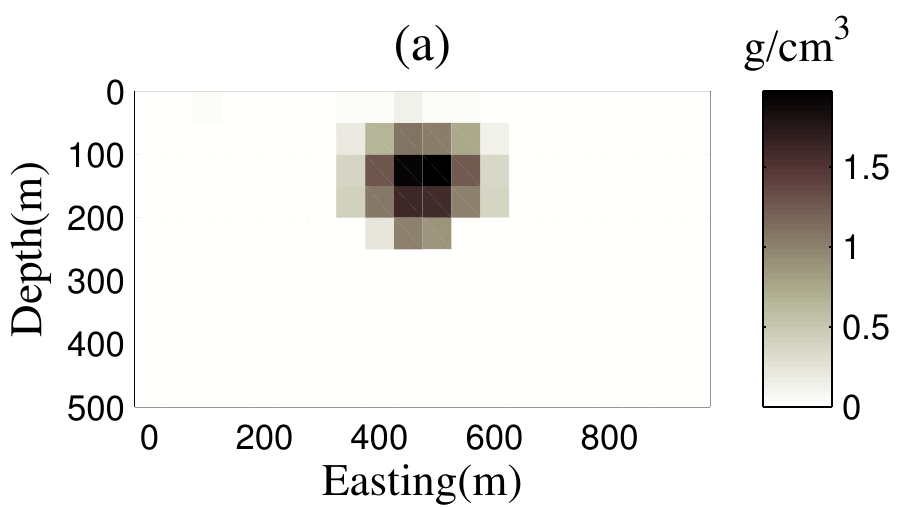}}
\subfigure{\label{5b}\includegraphics[width=.45\textwidth]{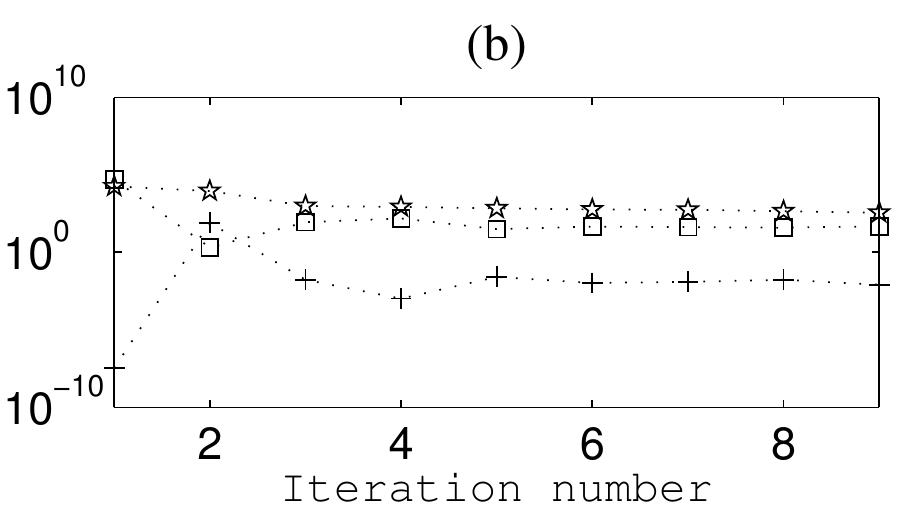}}
\subfigure{\label{5c}\includegraphics[width=.45\textwidth]{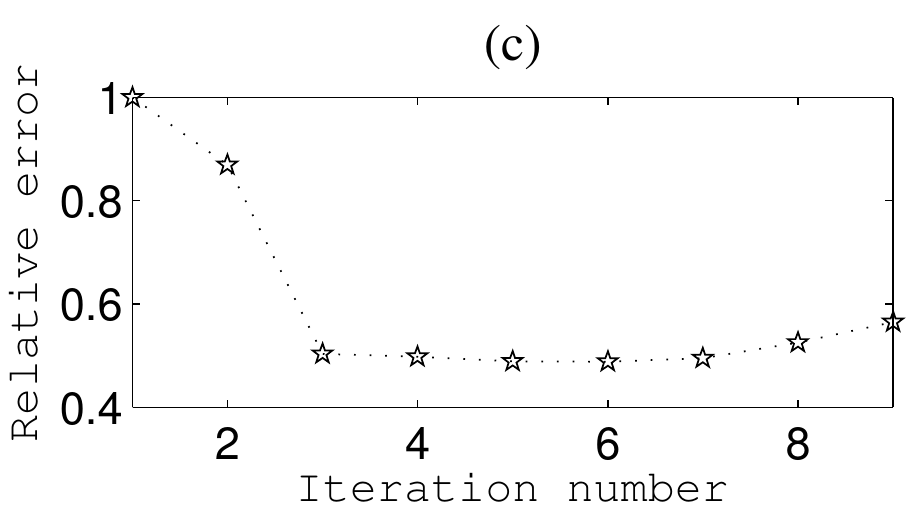}}
\subfigure{\label{5d}\includegraphics[width=.45\textwidth]{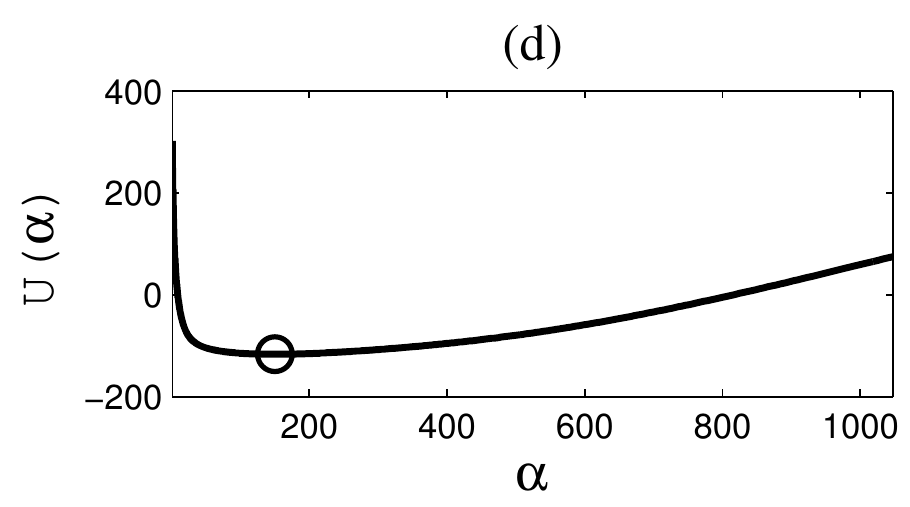}}
\caption {Contrasting constraint bounds  with  $\rho_{\mathrm{max}}=2$~g~cm$^{-3}$ as compared to $\rho_{\mathrm{max}}=1$~g~cm$^{-3}$ in Fig.~\ref{fig3}. UPRE is given for $k=4$.} \label{fig5}
\end{figure*}

To analyse the dependence of the Algorithm~\ref{svdalgorithm} on the density bounds, we select an unrealistic upper bound on the density,  $\rho_{\mathrm{max}}=2$~g~cm$^{-3}$.  All other parameters are chosen as before. Fig.~\ref{fig5} shows the inversion results. As compared with the results shown in Fig.~\ref{fig3}, the relative error and number of required iterations increases in this case. The reconstructed model has density values near to $2$~g~cm$^{-3}$ in the center, decreasing to values near $1$~g~cm$^{-3}$ at the border. This indicates that while the  perspective of the model is close to the original model, the knowledge of accurate bounds is required in reconstructing feasible models.  This is realistic for many geophysical investigations. 

\begin{figure*}
\subfigure{\label{6a}\includegraphics[width=.45\textwidth]{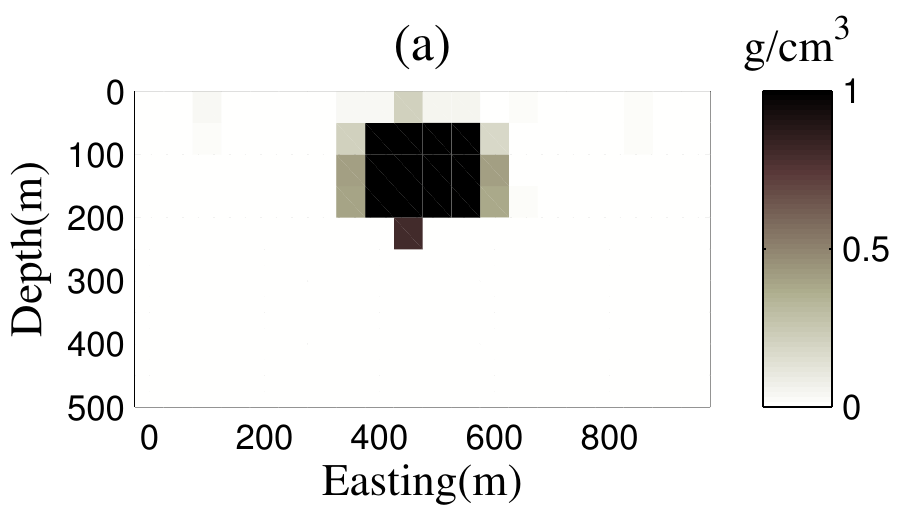}}
\subfigure{\label{6b}\includegraphics[width=.45\textwidth]{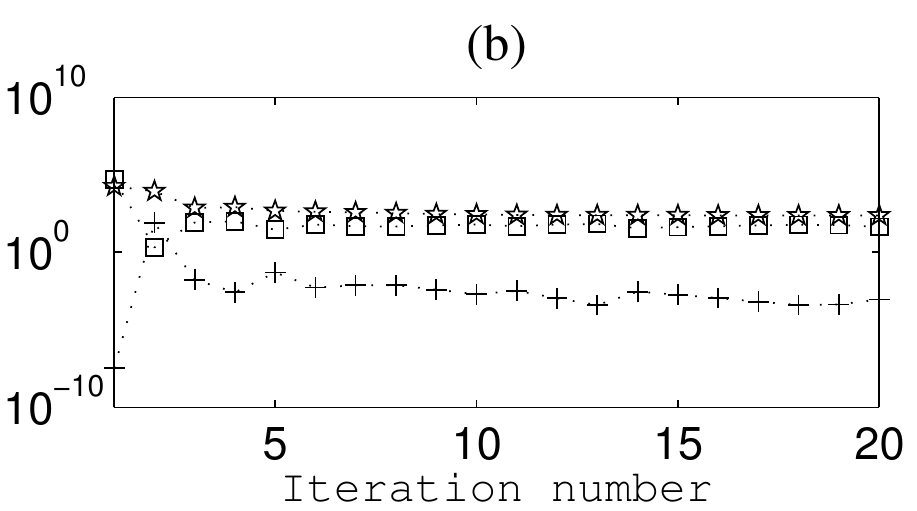}}
\subfigure{\label{6c}\includegraphics[width=.45\textwidth]{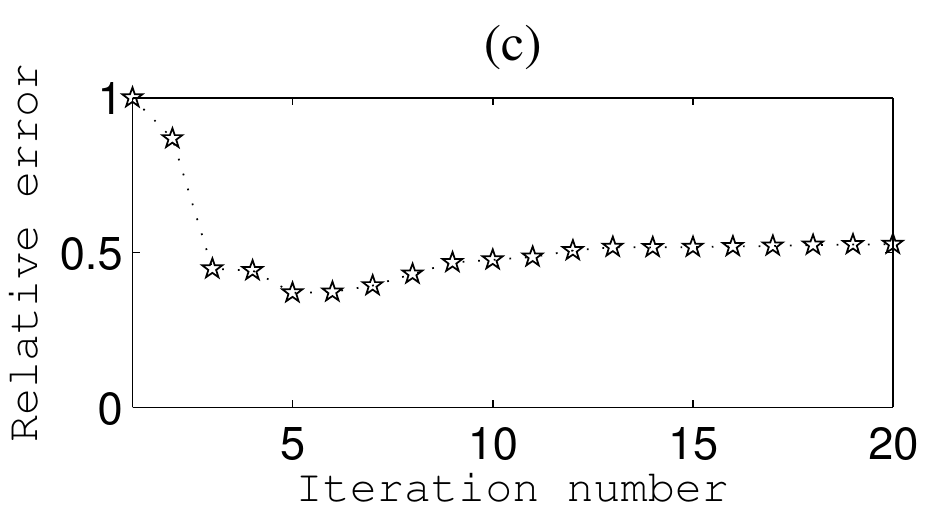}}
\subfigure{\label{6d}\includegraphics[width=.45\textwidth]{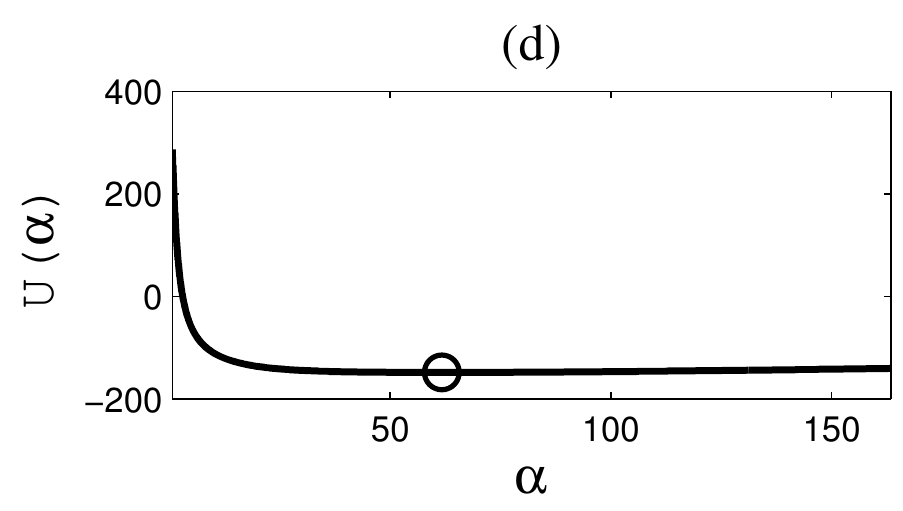}}
\caption {Contrasting results obtained out to $k=K_{\mathrm{max}}=20$, as compared to termination based on $\chi^2$ test in Fig.~\ref{fig3}. UPRE is given for $k=10$.} \label{fig6}
\end{figure*}

Finally, we examine Algorithm~\ref{svdalgorithm} without termination due to the $\chi^2$ test for convergence. We iterate out to  $K_{\mathrm{max}}=20$ to check the process of the inversion and regularization parameter estimation. The results are presented in Fig.~\ref{fig6}. The model is more focused but acceptable, and the regularization parameter converges to a fixed value.

\subsubsection{Degree of ill-conditioning} \label{degreeillposedness}
\begin{figure*}
\subfigure{\label{7a}\includegraphics[width=.3\textwidth]{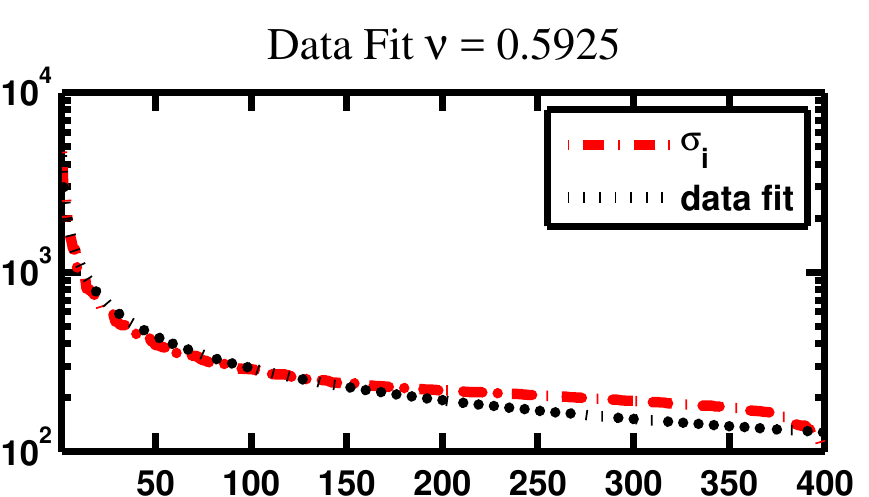}}
\subfigure{\label{7b}\includegraphics[width=.3\textwidth]{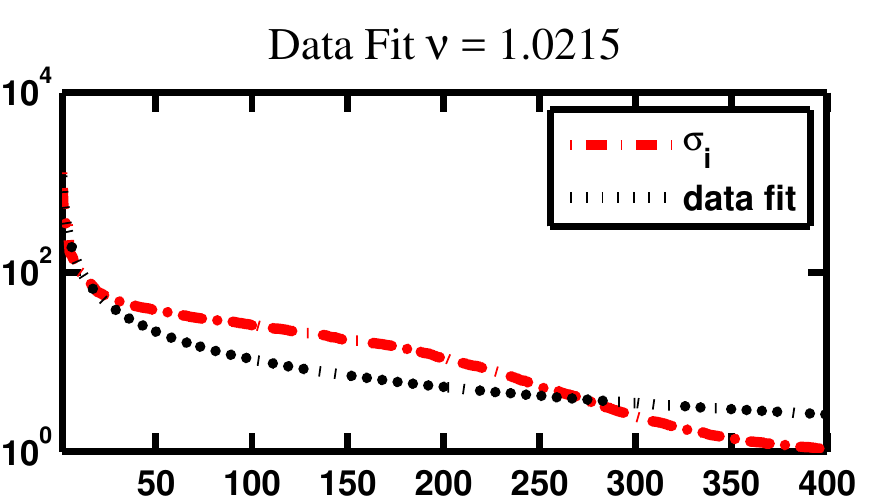}}
\subfigure{\label{7c}\includegraphics[width=.3\textwidth]{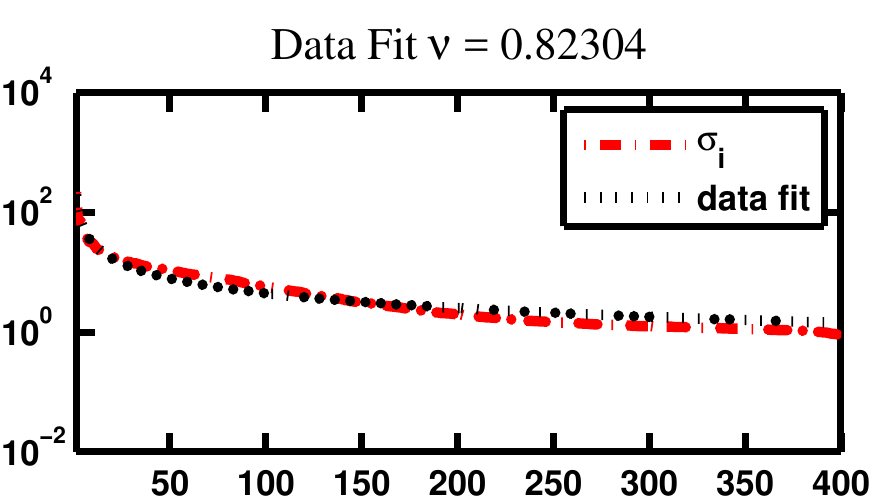}}
\caption {The original spectrum of matrix $\tilde{\tilde{G}}$ and the data fit function for a case $N2$ at the iteration $1$, $3$ and $5$ shown left to right with a title indicating $\nu$ at each iteration.} \label{fig7}
\end{figure*}

Before presenting the results using Algorithm~\ref{projectedalgorithm}, it is necessary to examine the underlying properties of the matrix $\tilde{\tilde{G}}$ in \refeq[globalfunctionh]. There are various notions of the degree to which a given property is ill-conditioned. We use the heuristic that if there exists $\nu$ such that  $\sigma_j \approx \mathcal{O}(j^{-\nu})$ then the system is mildly ill-conditioned for $0<\nu<1$ and moderately ill-conditioned for $\nu>1$. If $\sigma_j \approx \mathcal{O}(e^{-\nu j})$  for $\nu>1$ then the problem is said to be severely ill-conditioned,  see e.g. \cite{Hoff:86,HuJi:16}. These relationships relate to the speed with which the spectral values decay to $0$, decreasing much more quickly for the more severely ill-conditioned cases. Examination of the spectrum of $\tilde{\tilde{G}}$ immediately suggests that the problem is  mildly to moderately ill-conditioned, and under that assumption it is not difficult to do a nonlinear fit of the $\sigma_j$ to $C j^{-\nu}$ to find $\nu$. We note that the constant $C$ is irrelevant in determining the condition of the problem, which depends only on the ratio $\sigma_1/\sigma_n$, independent of $C$. The data fitting result for the case $N2$ is shown in Fig.~\ref{fig7} over $3$ iterations of the IRLS for a given sample. There is a very good fit to the data in each case, and $\nu$ is clearly iteration dependent. These results are  representative of the results obtained using both $N1$ and $N3$. We conclude that the gravity sensitivity matrix as used here is mildly to moderately ill-conditioned. 

\subsubsection{Solution using Algorithm~\ref{projectedalgorithm}}\label{cubeAlgorithm2}
Algorithm~\ref{projectedalgorithm} is used to reconstruct the model for $3$ different values of $t$, $t=100$, $200$ and $400$, in order to examine the impact of the size of the projected subspace  on the solution and the estimated parameter $\zeta$.  Here, in order to compare the algorithms,  the initial regularization parameter, $\zeta^{(1)}$, is set to the value that would be used on the full space. The results for cases $t=100$ and $200$ are given in Table~\ref{tab2new}.  Generally, for small $t$ the estimated  regularization parameter is less than the counterpart obtained for the  full case for the specific noise level. Comparing Tables~\ref{tab1} and ~\ref{tab2new} it is clear that with increasing $t$, the estimated $\zeta$ increases. For $t=100$ the results are not satisfactory.  The relative error is very large and the reconstructed model is generally not acceptable. Although the results with the least noise  are acceptable, they are still worse than those obtained with the other selected choices for $t$. In this case, $t=100$, and for high noise levels, the algorithm usually terminates when it reaches $k=K_{\mathrm{max}}=50$, indicating that the solution does not satisfy the noise level constraint. For $t=200$ the results are acceptable, although less satisfactory than the results obtained with the full space. With increasing $t$ the results improve, until for $t=m=400$ the results, not presented here,  are exactly those obtained with Algorithm~\ref{svdalgorithm}. This confirms the results in Renaut et al. \shortcite{RVA:2017} that when $t$ approximates the numerical rank of the sensitivity matrix, $ \zeta_{\mathrm{proj}}= \alpha_{\mathrm{full}}$.

\begin{table}
\begin{center}
\caption{The inversion results, for final regularization parameter $\zeta^{(K)}$, relative error $RE^{(K)}$ and number of iterations $K$ obtained by inverting the data from the cube using Algorithm~\ref{projectedalgorithm},  with $\epsilon^2=1e{-9}$,  and $\zeta^{(1)} = \alpha^{(1)}$ for the specific noise level as given in Table~\ref{tab1}. }\label{tab2new}
\begin{tabular}{c  c  c  c  c  c c c c c c c c }
\hline
&\multicolumn{3}{c}{t=100}&\multicolumn{3}{c}{t=200}\\ \hline 
&$\zeta^{(K)}$& $RE^{(K)}$& $K$ & $\zeta^{(K)}$& $RE^{(K)}$& $K$ \\ \hline 
$N1$ & 98.9(12.0)& 0.452(.043)& 10.0(0.7)& 102.2(11.3)& .329(.019)& 8.8(0.4) \\ \hline 
$N2$& 42.8(10.4)& 1.009(.184)& 28.1(10.9)&   43.8(7.0)& .429(.053)& 6.7(0.8) \\ \hline 
$N3$& 8.4(13.3)& 1.118(.108)& 42.6(15.6)&    27.2(6.3)& .463(.036)& 5.5(0.5)\\ \hline 
\end{tabular}
\end{center}
\end{table}

\begin{figure*}
\subfigure{\label{8a}\includegraphics[width=.45\textwidth]{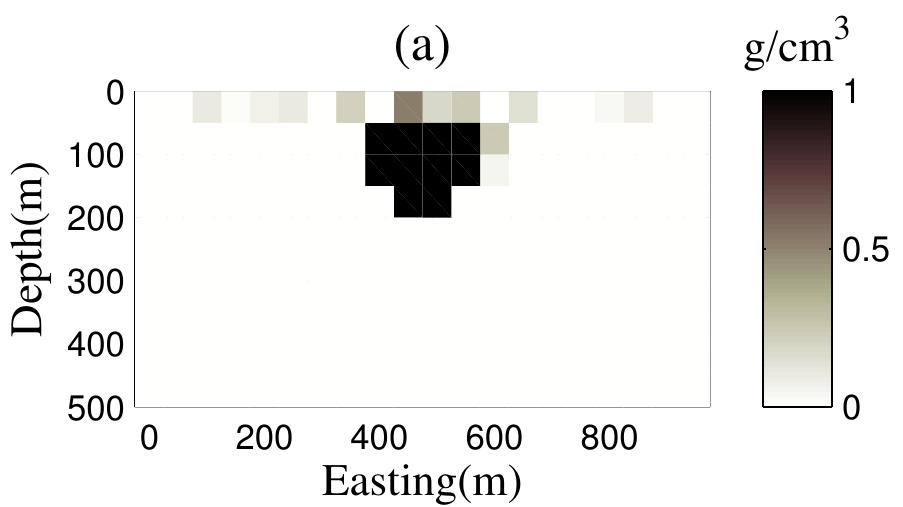}}
\subfigure{\label{8b}\includegraphics[width=.45\textwidth]{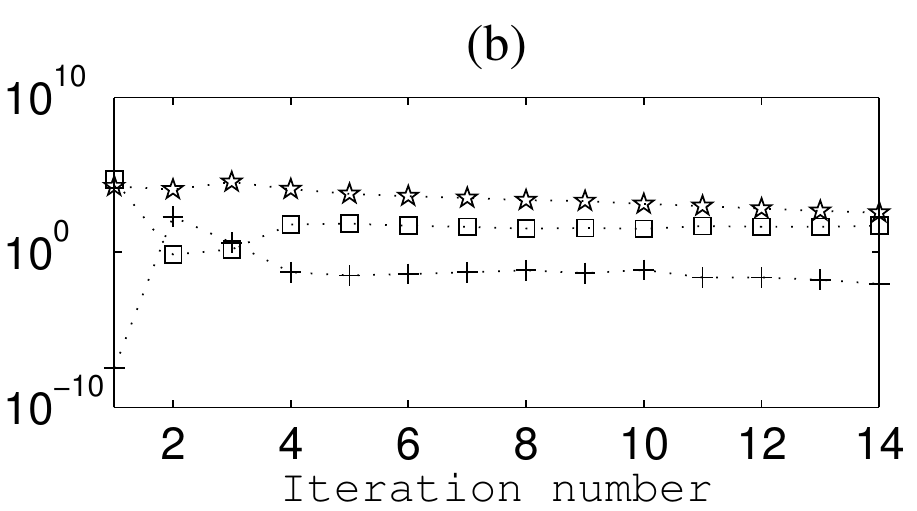}}
\subfigure{\label{8c}\includegraphics[width=.45\textwidth]{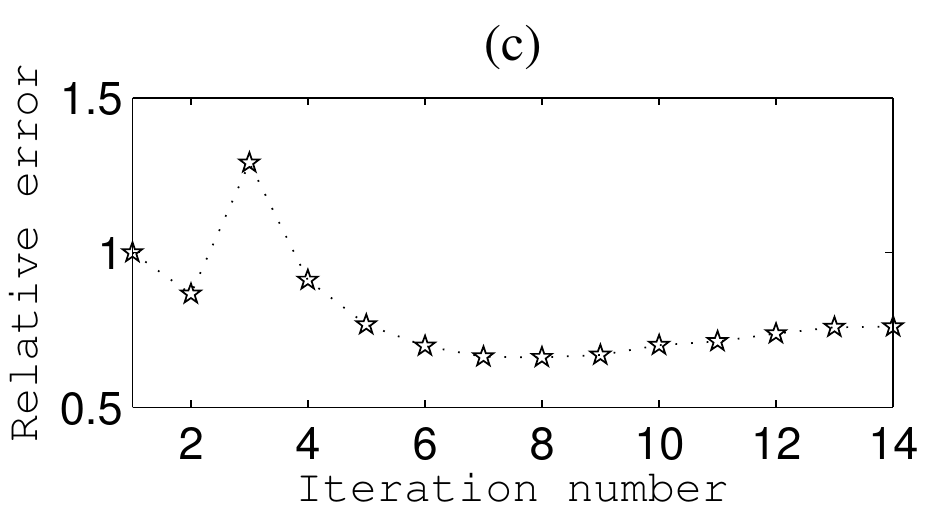}}
\subfigure{\label{8d}\includegraphics[width=.45\textwidth]{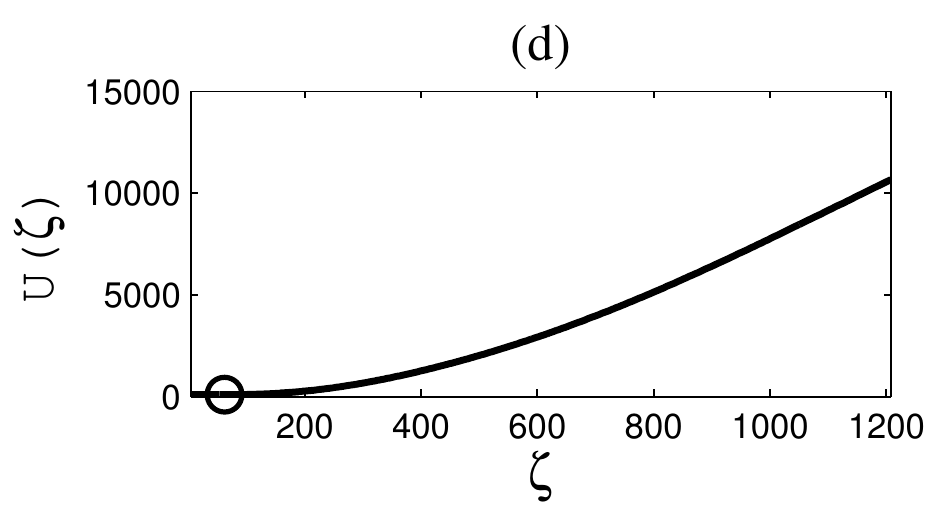}}
\caption {Inversion results using Algorithm~\ref{projectedalgorithm}  with $\epsilon^2=1e{-9}$ and $t=100$. UPRE is given for $k=4$.} \label{fig8}
\end{figure*}

\begin{figure*}
\subfigure{\label{9a}\includegraphics[width=.45\textwidth]{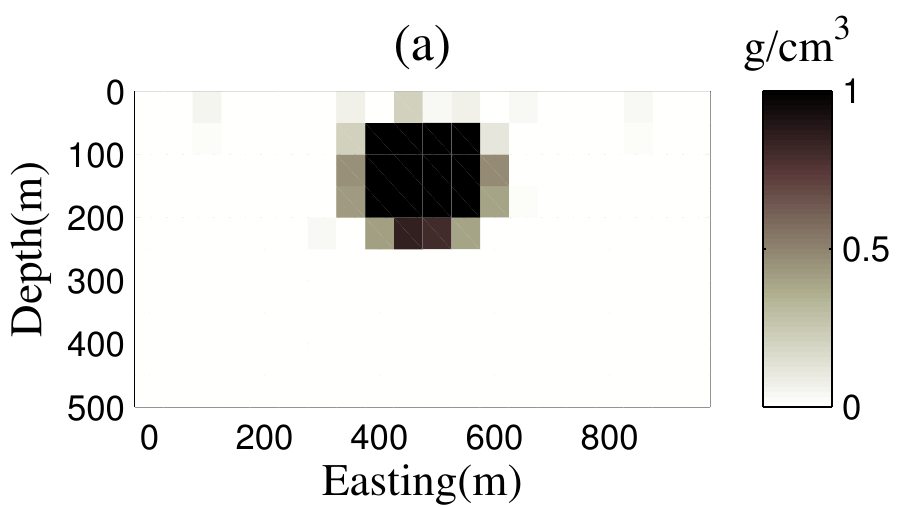}}
\subfigure{\label{9b}\includegraphics[width=.45\textwidth]{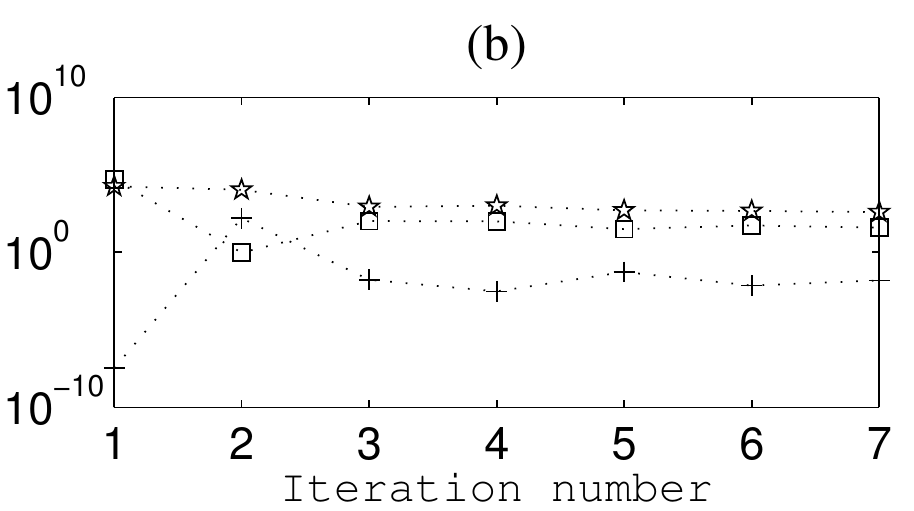}}
\subfigure{\label{9c}\includegraphics[width=.45\textwidth]{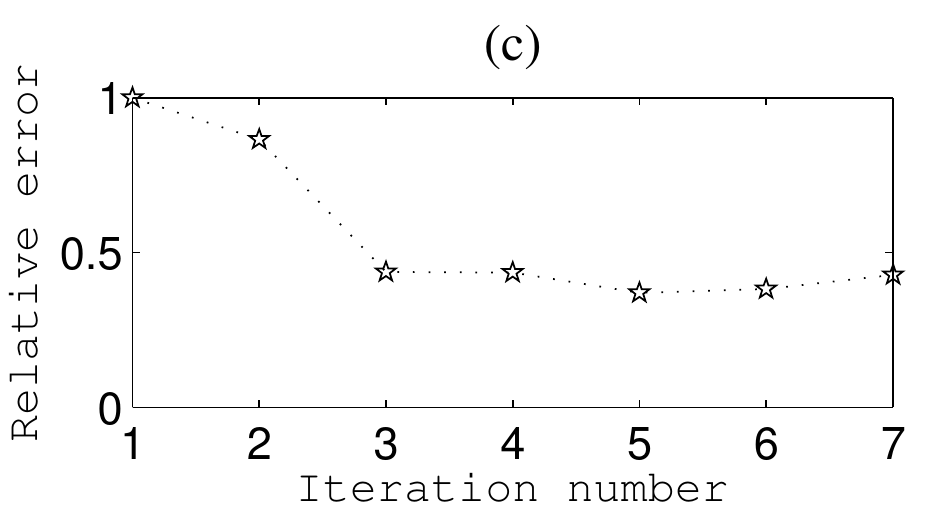}}
\subfigure{\label{9d}\includegraphics[width=.45\textwidth]{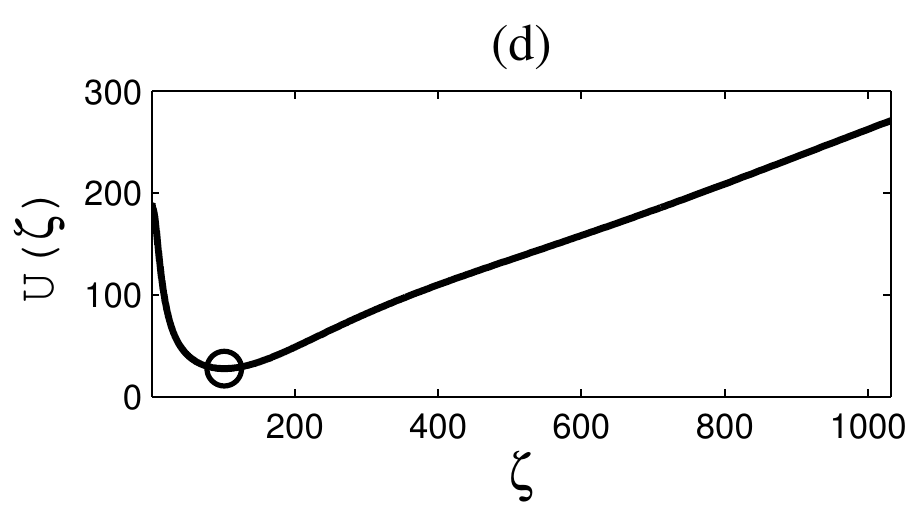}}
\caption {Inversion results using Algorithm~\ref{projectedalgorithm}  with $\epsilon^2=1e{-9}$ and $t=200$. UPRE is given for $k=4$.} \label{fig9}
\end{figure*}

An example case with noise level $N2$, for $t=100$ and $t=200$ is illustrated in Figs.~\ref{fig8} and \ref{fig9}. The reconstructed model for $t=200$ is acceptable, while for $t=100$ the results are completely wrong. For some right-hand sides $c$ with $t=100$, the reconstructed models may be much worse than shown in Fig.~\ref{8a}. For  $t=100$ and for high noise levels, usually the estimated value for $\zeta^{(k)}$ using  \refeq[subupre] for $1<k<K$ is small, corresponding to under regularization and yielding  a large error in the solution. To understand why the UPRE leads to under regularization  we illustrate the UPRE curves for iteration $k=4$ in Figs.~\ref{8d} and \ref{9d}. It is immediate that when using  moderately sized   $t$, $U(\zeta)$ may not have a well-defined minimum, because $U(\zeta)$ becomes increasingly flat.  Thus the algorithm may find a  minimum at a small regularization parameter which leads to under regularization of the higher index terms in the expansion, those for which the spectrum is not accurately captured. This can cause problems for moderate $t$,  $ t<200$. On the other hand, as $t$ increases, e.g. for $t=200$,  it appears that there is a unique minimum of $U(\zeta)$ and the regularization parameter found is appropriate.  Unfortunately, this situation creates a conflict with the need to use $t \ll m$ for large  problems. 

\subsubsection{Solution using algorithm~\ref{projectedalgorithm} and truncated UPRE}\label{cubeAlgorithm2plusTupre}
\begin{figure*}
\subfigure{\label{10a}\includegraphics[width=.4\textwidth]{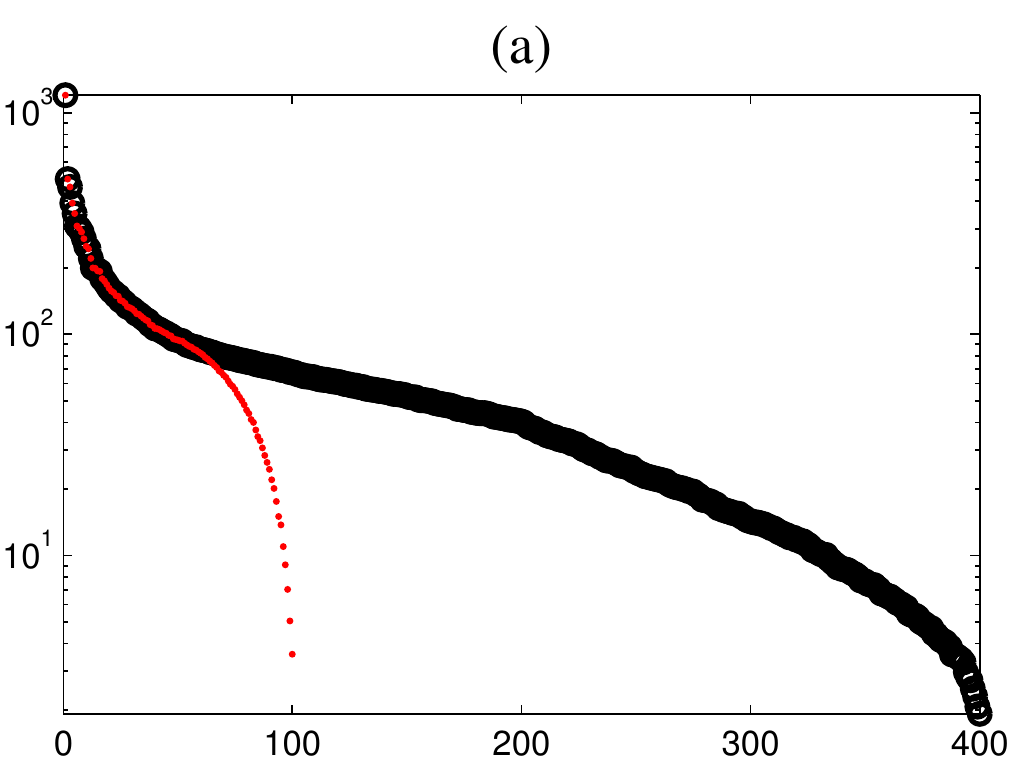}}
\subfigure{\label{10b}\includegraphics[width=.4\textwidth]{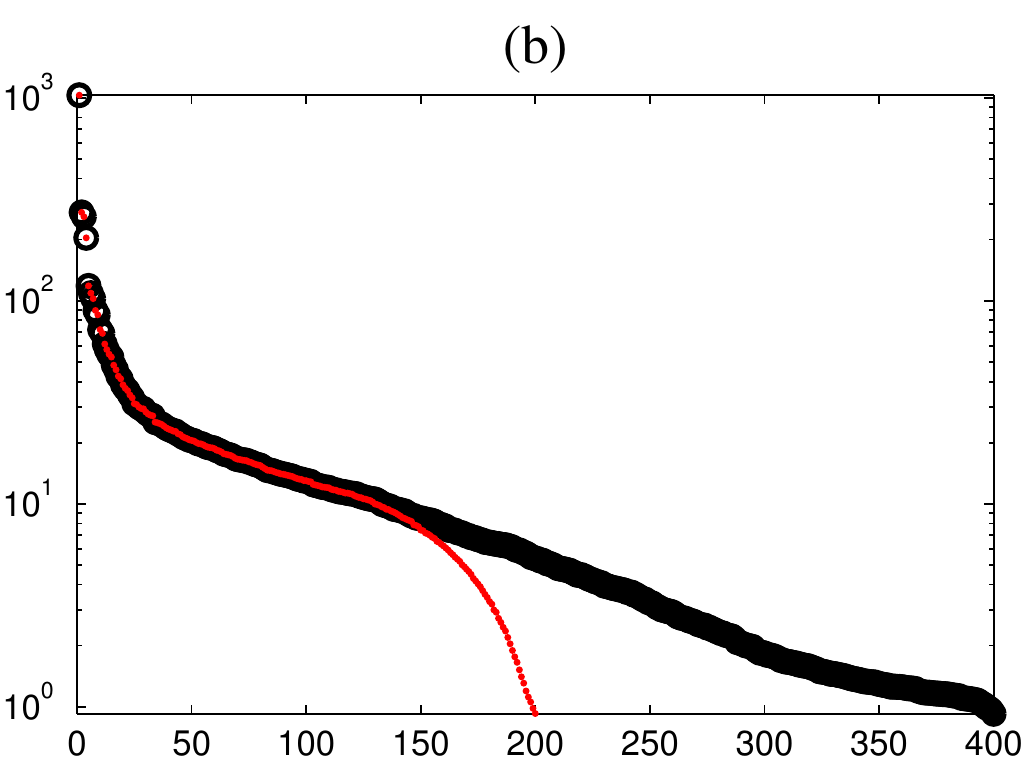}}
\caption {The singular values of original matrix $\tilde{\tilde{G}}$, block $\circ$, and  projected matrix $B_t$, red $\cdot$, at iteration $4$  using Algorithm~\ref{projectedalgorithm} with $\epsilon^2=1e{-9}$. (a) $t=100$;  and (b) $t=200$.}\label{fig10}
\end{figure*}

To determine the reason for the difficulty with using $U(\zeta)$ to find an optimal $\zeta$ for small to moderate $t$, we illustrate the singular values for $B_t$ and  $\tilde{\tilde{G}}$ in Fig.~\ref{fig10}. Fig.~\ref{10a} shows that using $t=100$ we do not estimate the first $100$ singular values accurately, rather only about the first $80$ are given accurately. The spectrum decays too rapidly, because $B_t$ captures the ill conditioning of the full system matrix. For $t=200$ we capture about $160$ to $170$ singular values correctly, while for $t=400$, not presented here, all the singular values are captured. Note, the behavior is similar for all iterations.  Indeed, because  $\tilde{\tilde{G}}$ is predominantly mildly ill-conditioned over all iterations the associated matrices $B_t$ do not accurately capture a right subspace of size $t$. Specifically, Huang $\&$ Jia \shortcite{HuJi:16} have shown  that the LSQR iteration captures the underlying Krylov subspace of the system matrix better when that matrix is severely or moderately ill-conditioned, and therefore LSQR has  better regularizing properties in these cases. On the other hand, for the mild cases LSQR is not sufficiently regularizing, additional regularization is needed, and the right singular subspace is contaminated by inaccurate spectral information that causes difficulty for effectively regularizing the projected problem in relation to the full problem. Thus using all the singular values from $B_t$  generates regularization parameters which are determined by the smallest singular values, rather than the dominant terms.

\begin{figure*}
\subfigure{\label{11a}\includegraphics[width=.45\textwidth]{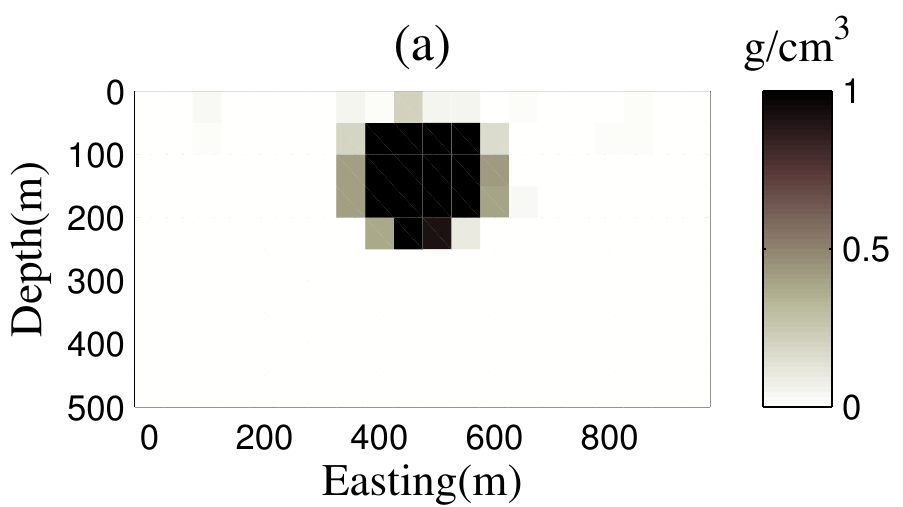}}
\subfigure{\label{11b}\includegraphics[width=.45\textwidth]{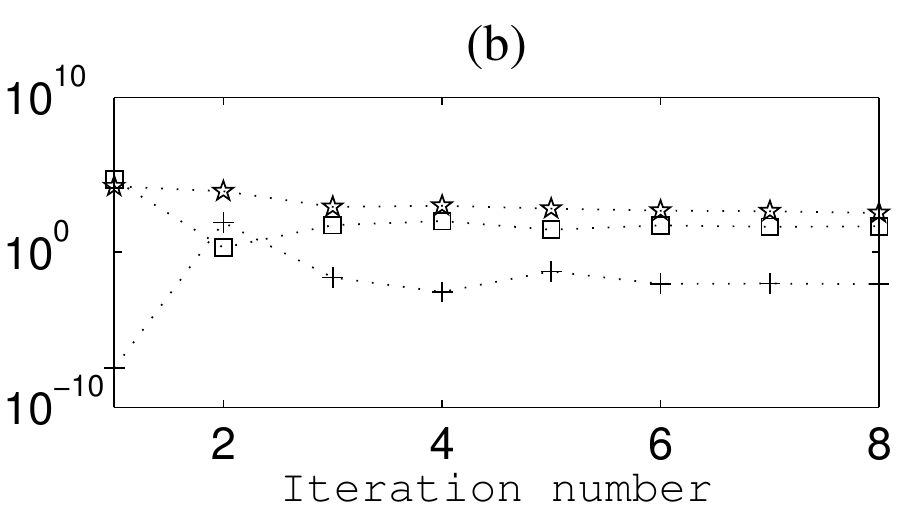}}
\subfigure{\label{11c}\includegraphics[width=.45\textwidth]{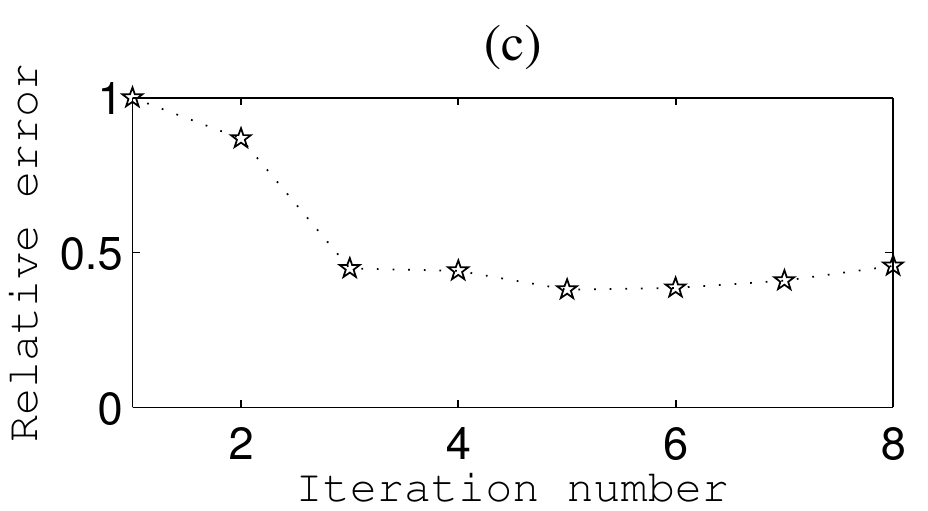}}
\subfigure{\label{11d}\includegraphics[width=.45\textwidth]{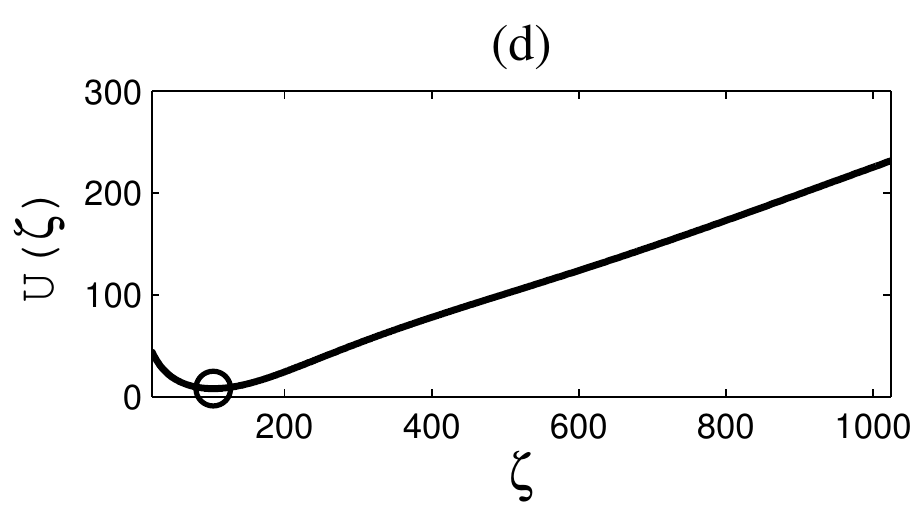}}
\caption {Reconstruction using Algorithm~\ref{projectedalgorithm}, with $\epsilon^2=1e{-9}$ and TUPRE  when $t=100$ is chosen. The TUPRE function at iteration $4$.} \label{fig11}
\end{figure*}

\begin{figure*}
\subfigure{\label{12a}\includegraphics[width=.45\textwidth]{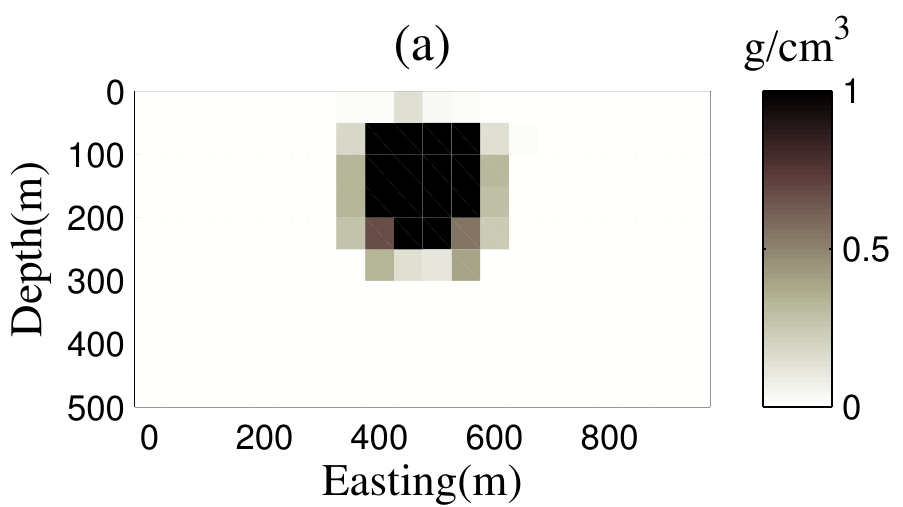}}
\subfigure{\label{12b}\includegraphics[width=.45\textwidth]{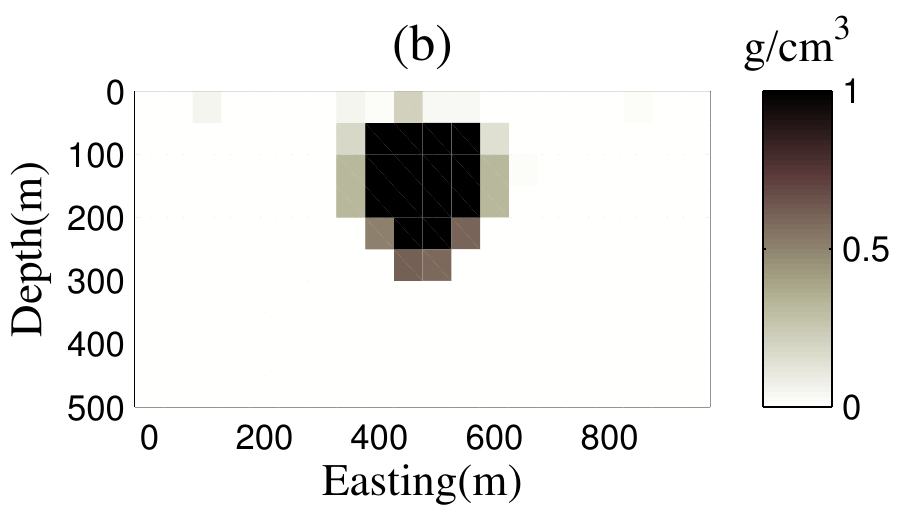}}
\subfigure{\label{12c}\includegraphics[width=.45\textwidth]{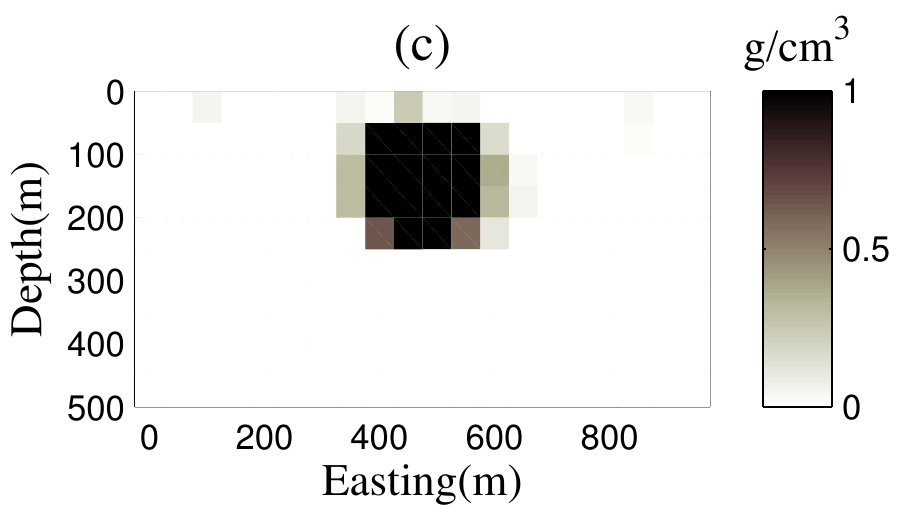}}
\subfigure{\label{12d}\includegraphics[width=.45\textwidth]{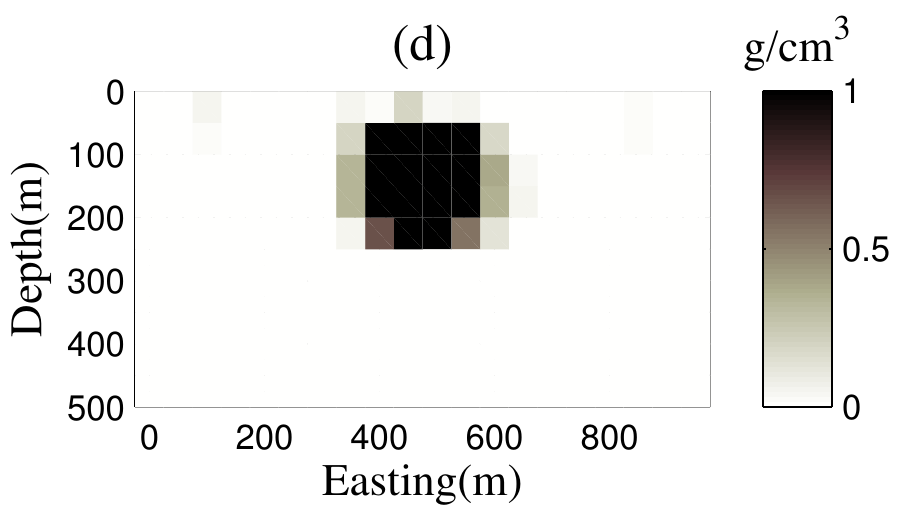}}
\caption {Reconstruction using Algorithm~\ref{projectedalgorithm}, with $\epsilon^2=1e{-9}$ and  TUPRE. (a) $t=10$; (b) $t=20$;  (c)  $t=30$ and (d)  $t=40$.} \label{fig12}
\end{figure*}

Now  suppose that we use $t$ steps of the GKB on matrix $\tilde{\tilde{G}}$ to obtain $B_t$, but use 
\begin{equation}\label{ttrunc}
t_{\mathrm{trunc}} = \omega \, t \quad \omega<1,
\end{equation}
singular values of $B_t$ in estimating  $\zeta$, in step \ref{stepzeta} of Algorithm~\ref{projectedalgorithm}. Our  examinations suggest taking   $ 0.7 \leq \omega < 1$. With this choice the smallest singular values of $B_t$ are ignored in estimating $\zeta$. We denote the approach, in which at step \ref{stepzeta} in Algorithm~\ref{projectedalgorithm}  we use truncated UPRE (TUPRE) for $t_{\mathrm{trunc}}$.  We comment that the approach may work equally well for alternative regularization techniques, but this is not a topic of the current investigation, neither is a detailed investigation for the choice of $\omega$. Furthermore, we note this  is not a standard filtered truncated SVD for the solution. The truncation of the spectrum is used in the estimation of the regularization parameter, but all $t$ terms of the singular value expression are used for the solution. To show the efficiency of TUPRE, we run the inversion algorithm for case $t=100$ for which the original results are not realistic.  The results using TUPRE are given in Table~\ref{tab3} for noise $N1$, $N2$ and $N3$, and illustrated, for right-hand side $c=7$ for noise $N2$, in Fig.~\ref{fig11}. Comparing the new results with those obtained in section~\ref{cubeAlgorithm2}, demonstrates that the TUPRE method yields a significant improvement in the solutions. Indeed, Fig.~\ref{11d} shows the  existence of a well-defined minimum for $U(\zeta)$   at iteration $k=4$, as compared to Fig.~\ref{8d}.  Further, it is possible to now obtain acceptable solutions with small $t$, which is demonstrated for reconstructed models obtained using $t=10$, $20$, $30$ and $40$ in Fig.~\ref{fig12}. All results are reasonable, hence  indicating that  the method can be used with high confidence even for small $t$. We note here that $t$ should be selected as small as possible, $t \ll m$, in order that the implementation of the GKB is fast, while simultaneously the dominant right singular subspace of the original matrix should be captured. In addition, for very small $t$, for example $t=10$ in Fig.~\ref{fig12}, the method chooses the smallest singular value as regularization parameter, there is no nice minimum for the curve. Our analysis suggests that $t > m/20$ is a suitable choice for application in the algorithm, although smaller $t$ may be used as $m$ increases. In all our test examples we use $\omega = 0.7$. 

\begin{table}
\begin{center}
\caption{The inversion results for final regularization parameter $\alpha^{(K)}$, relative error $RE^{(K)}$ and number of iterations $K$  obtained by inverting the data from the cube using Algorithm~\ref{projectedalgorithm}, with $\epsilon^2=1e{-9}$, using TUPRE with $t=100$, and $\zeta^{(1)} = \alpha^{(1)}$ for the specific noise level as given in Table~\ref{tab1}. }\label{tab3}
\begin{tabular}{c  c  c  c  c }
\hline
Noise&      $\zeta^{(K)}$& $RE^{(K)}$& $K$  \\ \hline
$N1$ &    131.1(9.2)& 0.308(0.007)& 6.7(0.8)\\
$N2$ &  52.4(6.2)& 0.422(0.049)& 6.8(0.9)\\
$N3$ &   30.8(3.3) & 0.483(0.060) & 6.9(1.1)\\ \hline
\end{tabular}
\end{center}
\end{table}

\subsection{Model of multiple embedded bodies}\label{multiplebodies}
\begin{figure*}
\subfigure{\label{13a}\includegraphics[width=.5\textwidth]{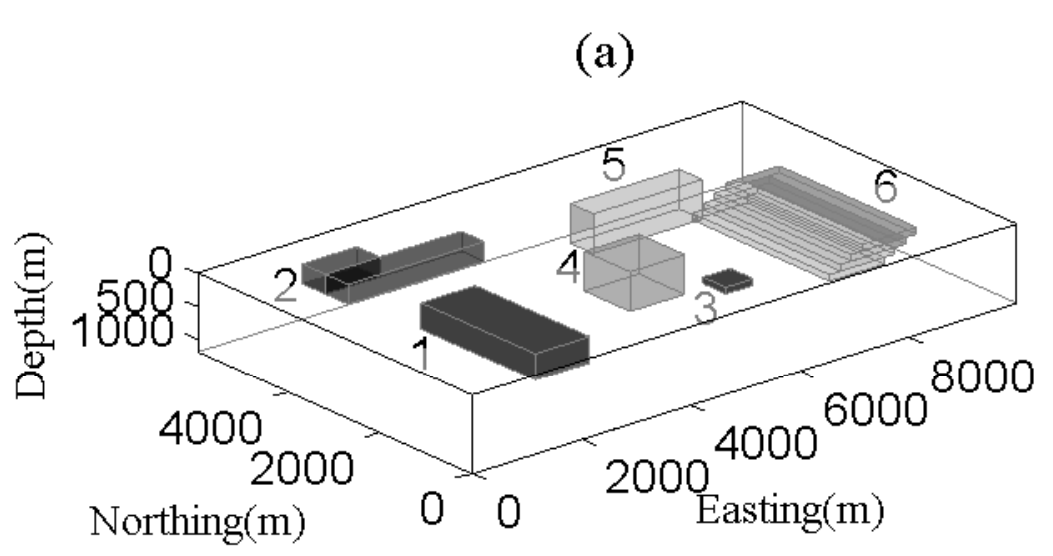}}
\subfigure{\label{13b}\includegraphics[width=.45\textwidth]{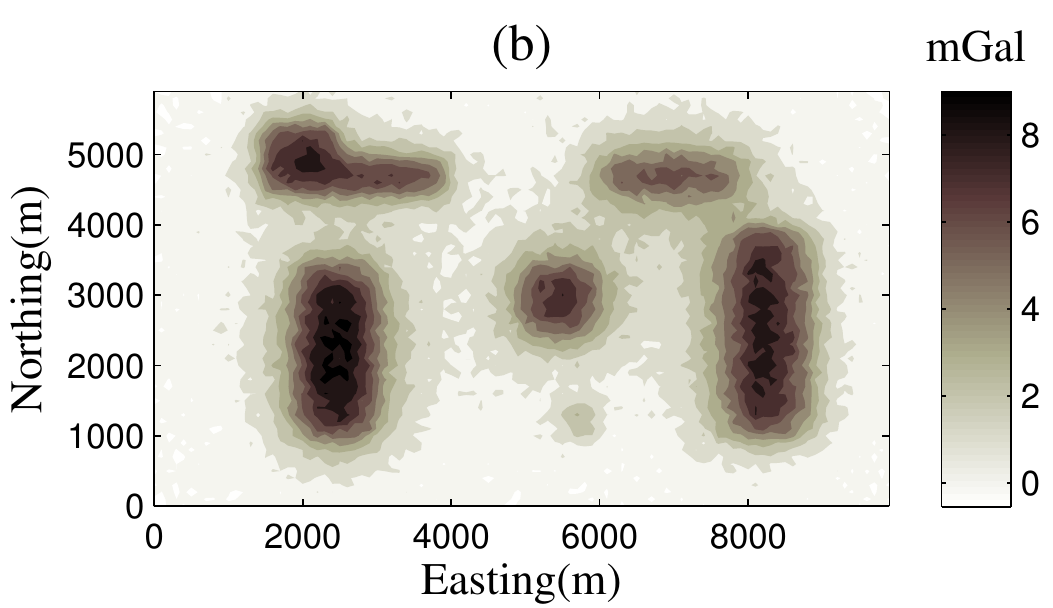}}
\caption {  (a) The perspective view of the model. Six different bodies embedded in an homogeneous background. Darker and brighter bodies have the densities $1$~g~cm$^{-3}$ and $0.8$~g~cm$^{-3}$, respectively; (b) The noise contaminated gravity anomaly due to the model.} \label{fig13}
\end{figure*}

\begin{figure*}
\subfigure{\label{14a}\includegraphics[width=.45\textwidth]{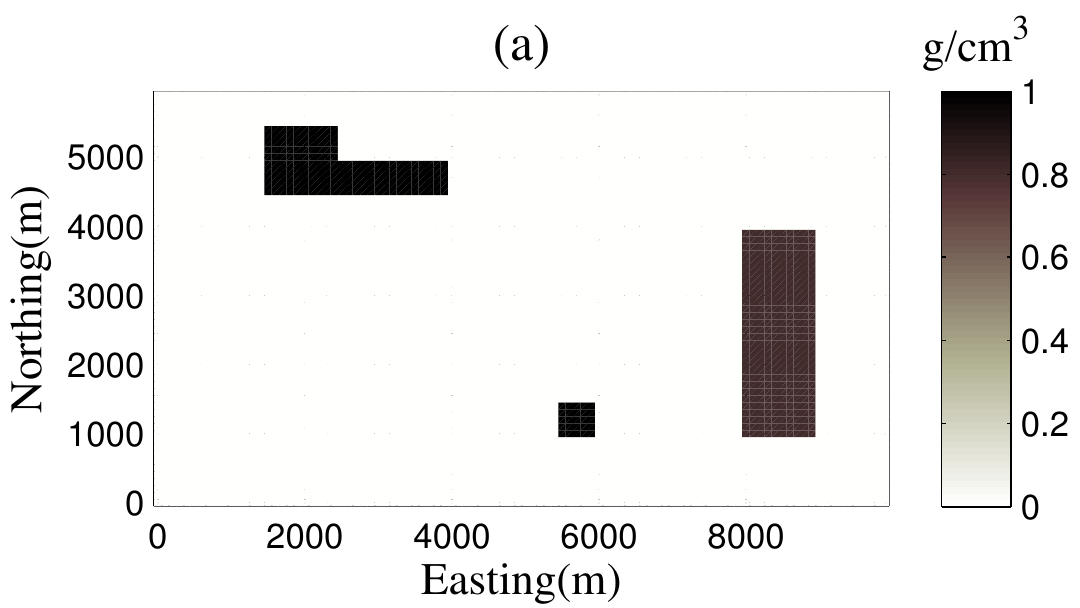}}
\subfigure{\label{14b}\includegraphics[width=.45\textwidth]{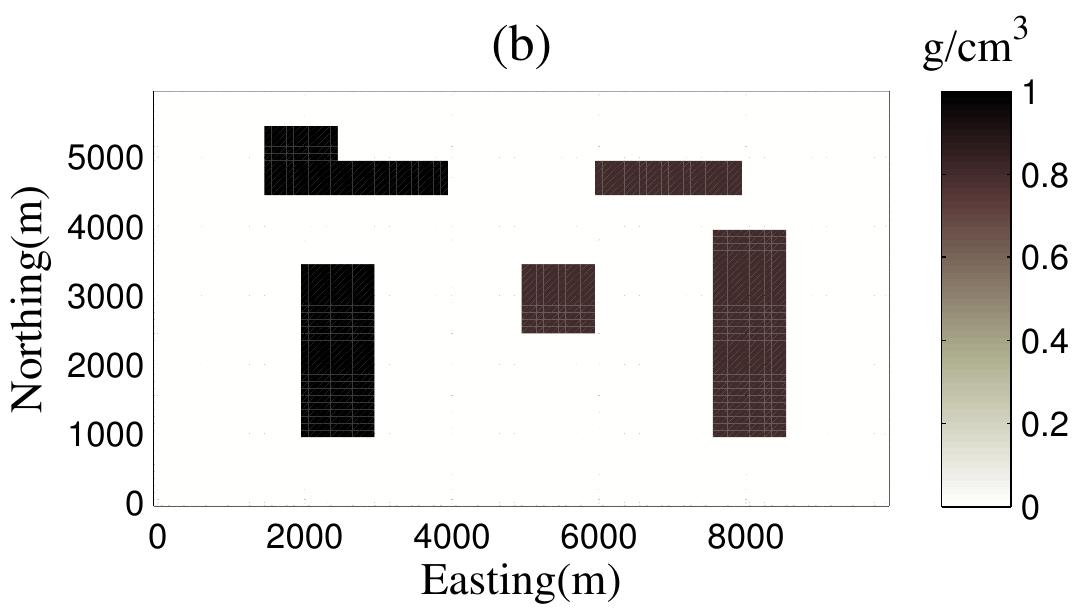}}
\subfigure{\label{14c}\includegraphics[width=.45\textwidth]{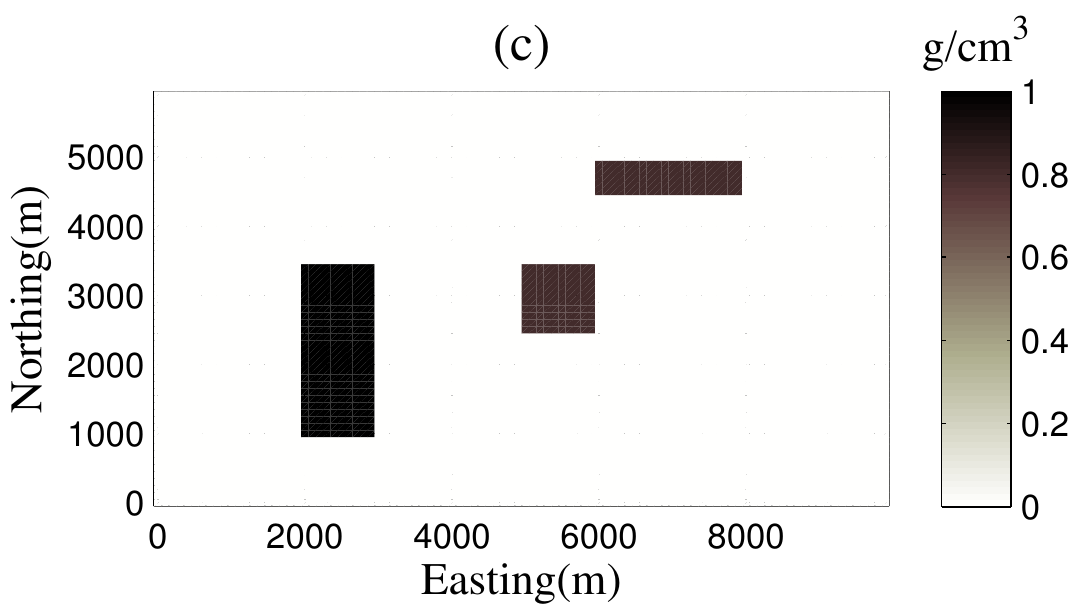}}
\subfigure{\label{14d}\includegraphics[width=.45\textwidth]{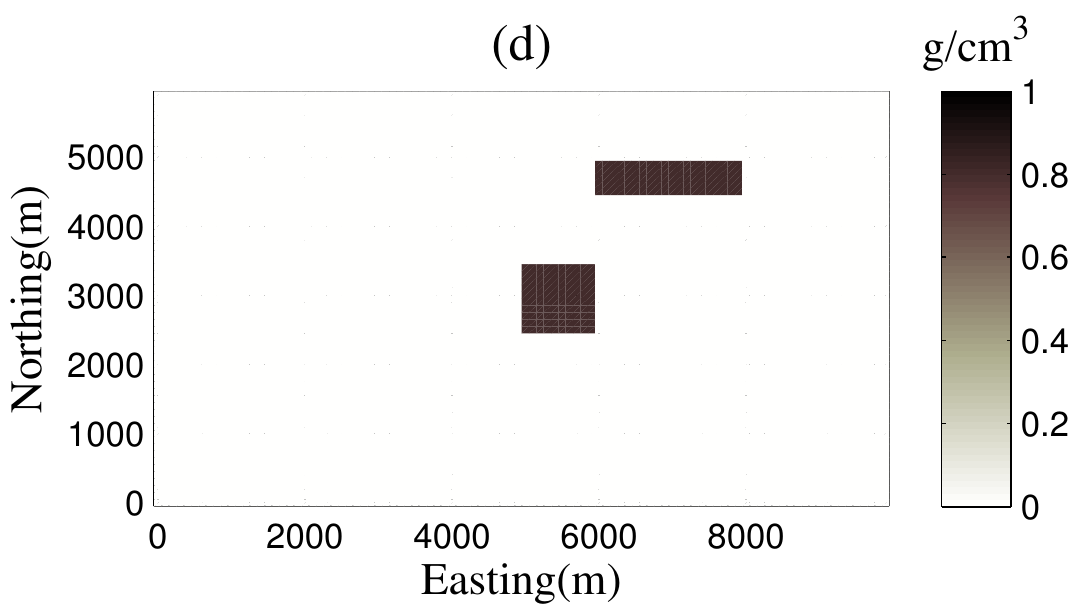}}
\caption{ The original model is displayed in four plane-sections. The depths of the sections are: (a) $Z=100$ m; (b) $Z=300$ m; (c) $Z=500$ m and (d)  $Z=700$ m.} \label{fig14}
\end{figure*}

A model consisting of six bodies with various geometries, sizes, depths and densities is used to verify the ability and limitations of  Algorithm~\ref{projectedalgorithm} implemented with TUPRE for the recovery of a larger and more complex structure. Fig.~\ref{13a} shows a perspective view of this model. The density and dimension of each body is given in Table~\ref{tab4}. Fig.~\ref{fig14} shows four plane-sections of the model. The surface gravity data are calculated on a $100 \times 60$ grid with $100$~m spacing, for a data vector of length $6000$. Noise is added to the exact data vector as in  \refeq[noisydata] with $(\tau_1, \tau_2)=(.02, .001)$. Fig.~\ref{13b} shows the noise-contaminated data. The subsurface extends to depth $1200$~m with cells of size $100$~m in each dimension yielding the unknown model parameters to be found on $ 100 \times 60 \times 12 = 72000$ cells. The inversion assumes $\bfma=\mathbf{0}$, $\beta=0.6$, $\epsilon^2=1e{-9}$ and imposes density bounds $\rho_{\mathrm{min}}=0$~g~cm$^{-3}$  and $ \rho_{\mathrm{max}}=1$~g~cm$^{-3}$. The iterations are terminated when $\chi_{\mathrm{Computed}}^2 \leq 6110 $,  or $k>K_{\mathrm{max}}=20$. The inversion is performed using Algorithm~\ref{projectedalgorithm} but with the TUPRE  parameter choice method  for step \ref{stepzeta}  with  $\omega=0.7$. The initial regularization parameter is  $\zeta^{(1)} = (n/m)^{3.5}(  \gamma_1/\mathrm{mean(\gamma_i)})$, for $\gamma_i$, $i=1:t $.

\begin{table}
\begin{center}
\caption{Assumed parameters for model with multiple bodies.}\label{tab4}
\begin{tabular}{c  c  c  c }
\hline
Source number&   $x\times y \times z$ dimensions (m) & Depth (m)& Density (g~cm$^{-3}$)  \\ \hline
1 &  $1000\times 2500 \times 400$  &    $200$     &   $1     $\\
2 &  $2500 \times 1000 \times 300$  &   $ 100$     & $   1 $    \\
3 &  $500 \times 500 \times 100$  &    $100    $ &    $ 1    $ \\
4 &  $1000 \times 1000 \times 600$  &    $200    $ &    $ 0.8    $ \\
5 &  $2000 \times 500 \times 600$  &    $200    $ &    $ 0.8    $ \\
6 &  $1000 \times 3000 \times 400$ &   $100      $&     $ 0.8 $    \\ \hline
\end{tabular}
\end{center}
\end{table}

\begin{figure*}
\subfigure{\label{15a}\includegraphics[width=.45\textwidth]{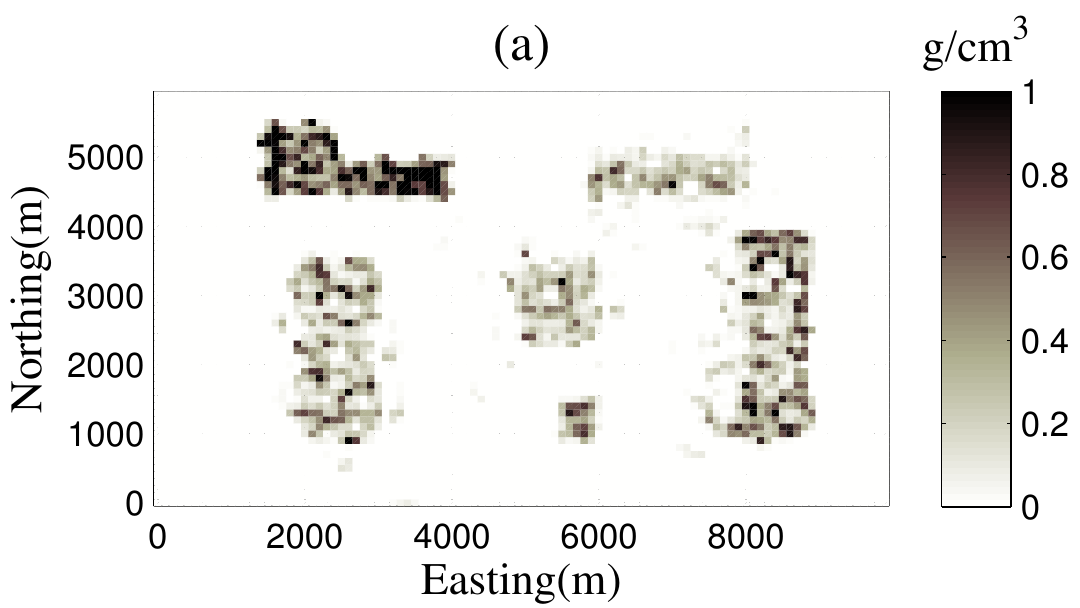}}
\subfigure{\label{15b}\includegraphics[width=.45\textwidth]{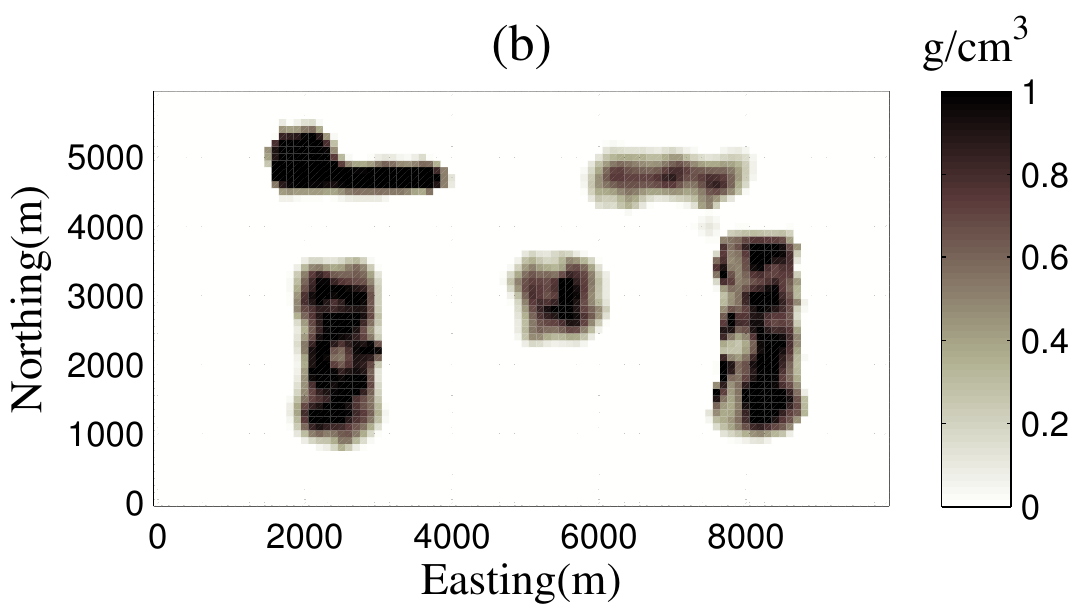}}
\subfigure{\label{15c}\includegraphics[width=.45\textwidth]{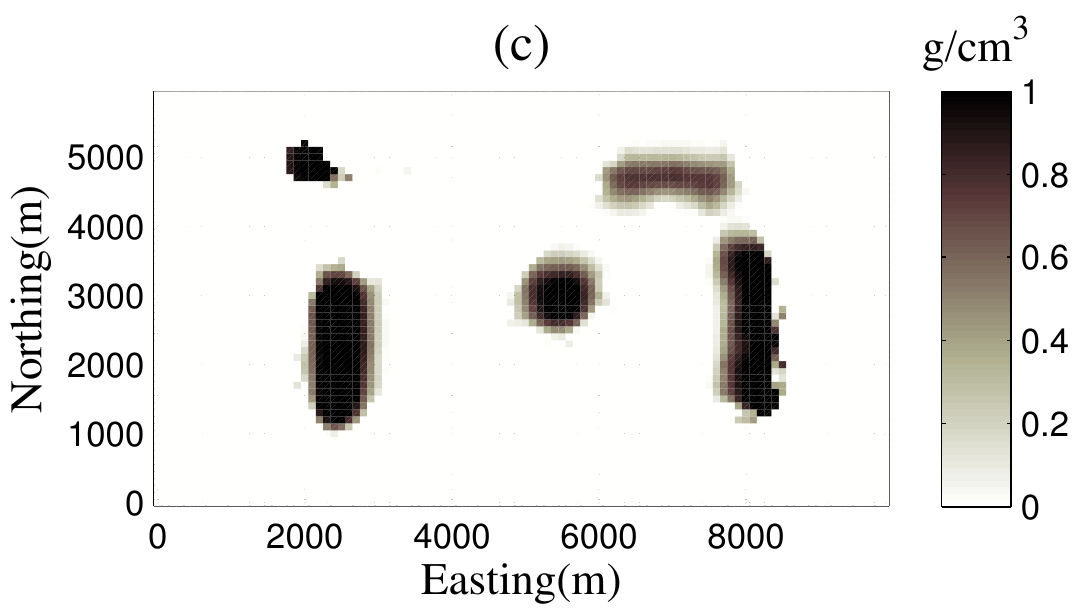}}
\subfigure{\label{15d}\includegraphics[width=.45\textwidth]{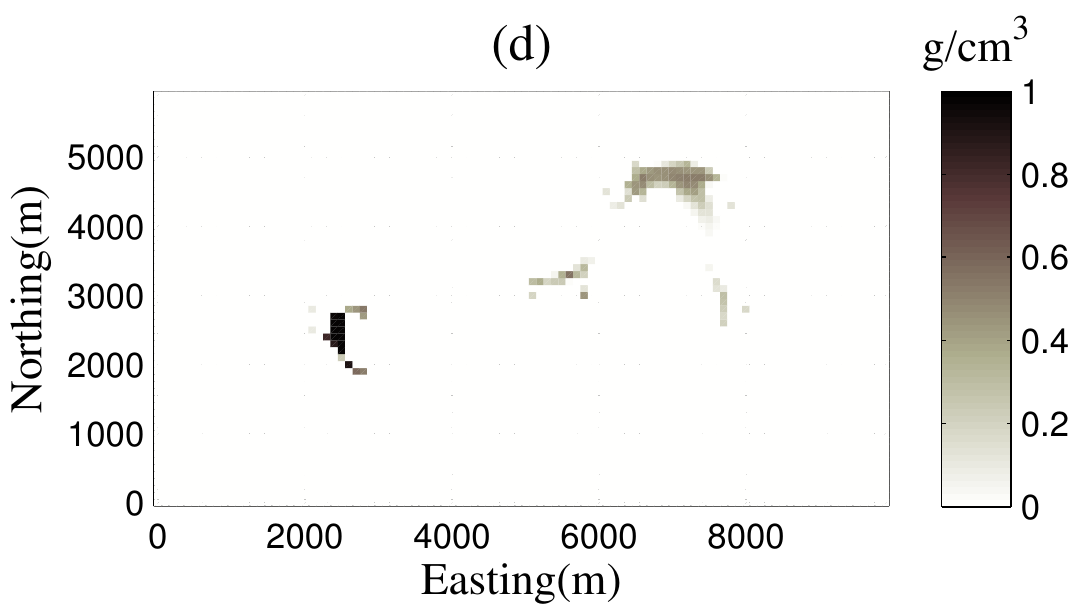}}
\caption{ For the data in Fig.~\ref{13b}: The reconstructed model  using Algorithm~\ref{projectedalgorithm} with $t=350$ and TUPRE method. The depths of the sections are: (a) $Z=100$ m; (b) $Z=300$ m; (c) $Z=500$ m and (d)  $Z=700$ m.} \label{fig15}
\end{figure*}

\begin{figure*}
\subfigure{\label{16a}\includegraphics[width=.4\textwidth]{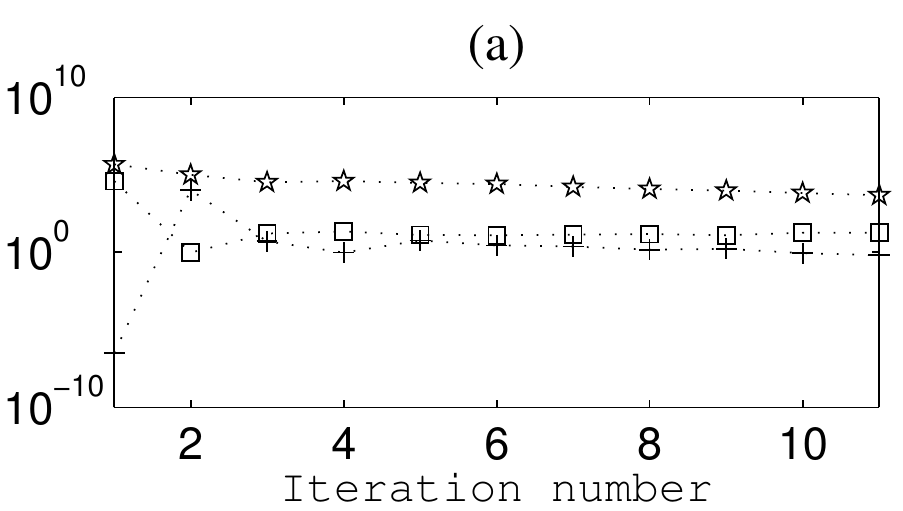}}
\subfigure{\label{16b}\includegraphics[width=.4\textwidth]{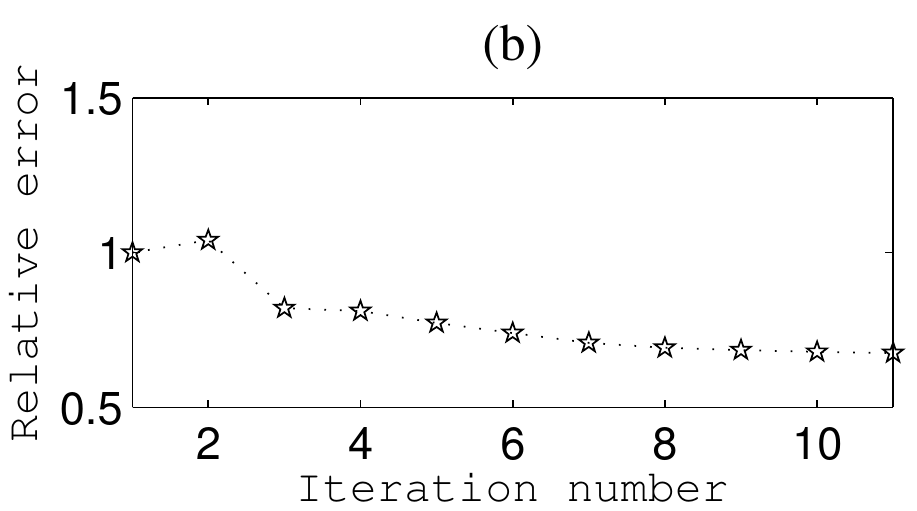}}
\subfigure{\label{16c}\includegraphics[width=.4\textwidth]{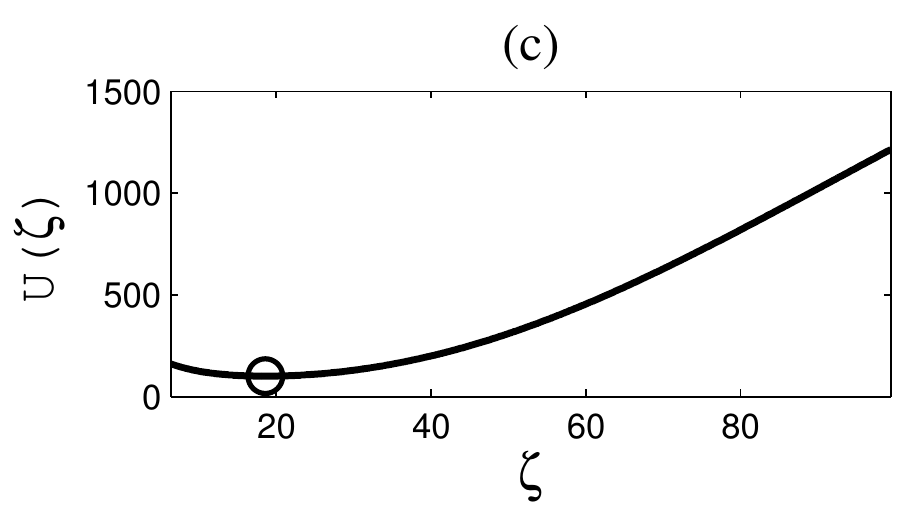}}
\caption {For the reconstructed model in Fig.~\ref{fig15}. In (c) the TUPRE at iteration $11$.} 
\label{fig16}
\end{figure*}

\begin{figure*}
\includegraphics[width=0.5\textwidth]{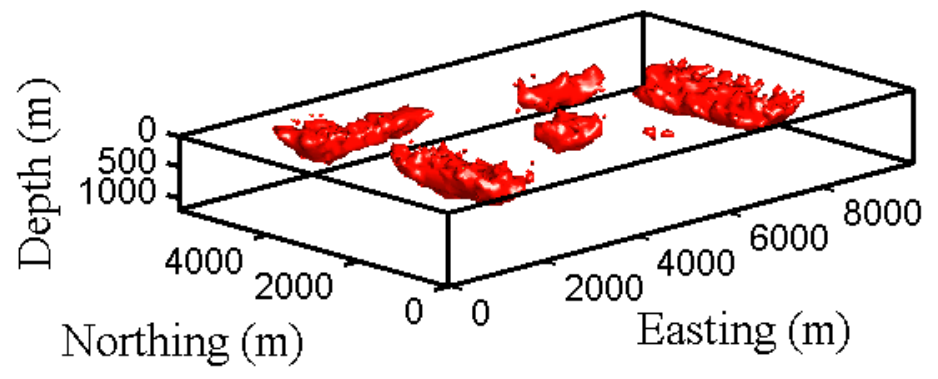}
\caption{The isosurface of the $3$-D inversion results with a density greater than  $0.6$~g~cm$^{-3}$ using Algorithm~\ref{projectedalgorithm} and TUPRE method  with $t=350$.}\label{fig17}
\end{figure*}

We illustrate the inversion results for $t=350$ in Figs.~\ref{fig15} and \ref{fig16}. The reconstructed model is close to the original model, indicating  that the algorithm is generally  able to give realistic results even for a complicated model. The inversion is completed on a Core i$7$ CPU $3.6$ GH desktop computer in nearly $30$ minutes. As illustrated, the horizontal borders of the bodies are recovered and the depths to the top  are close to those of  the original model. At the intermediate depth, the shapes of the anomalies are acceptably reconstructed, while deeper into the subsurface the model does not match the original so well.  Anomalies $1$ and $2$ extend to a greater depth than in the true case. In addition, Fig.~\ref{fig17} shows a $3$-D perspective view of the results for densities greater than $0.6$~g~cm$^{-3}$. We note here that comparable results can be obtained for smaller $t$, for example $t=50$, with a significantly reduced computational time, as is the case for the cube in Fig.~\ref{fig12}. We present the case for larger $t$ to demonstrate that the TUPRE algorithm is effective for finding $\zetaopt$ as $t$ increases.

\begin{figure*}
\subfigure{\label{18a}\includegraphics[width=.45\textwidth]{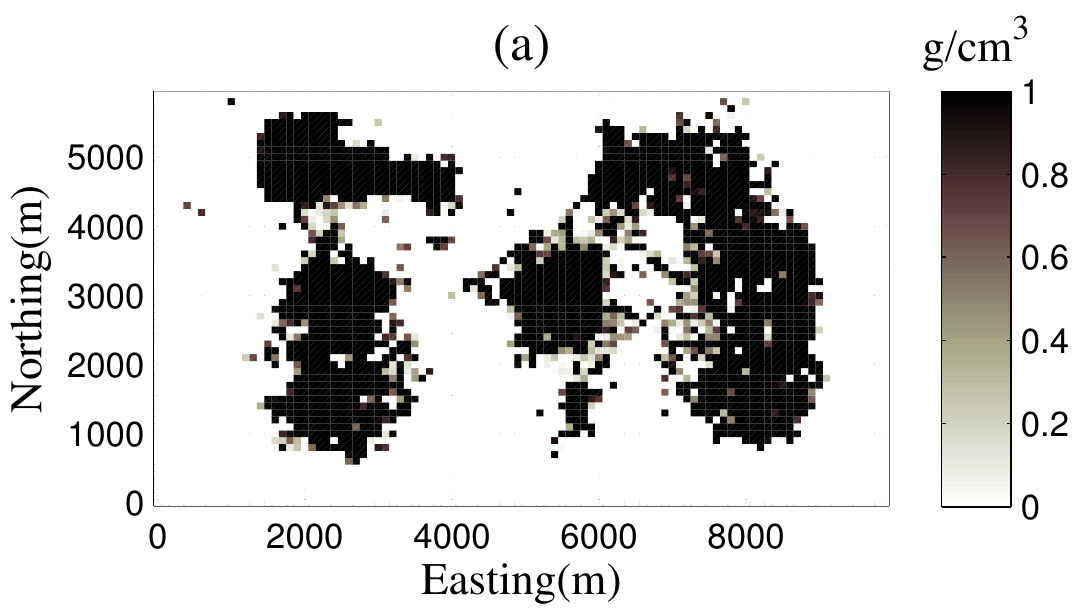}}
\subfigure{\label{18b}\includegraphics[width=.45\textwidth]{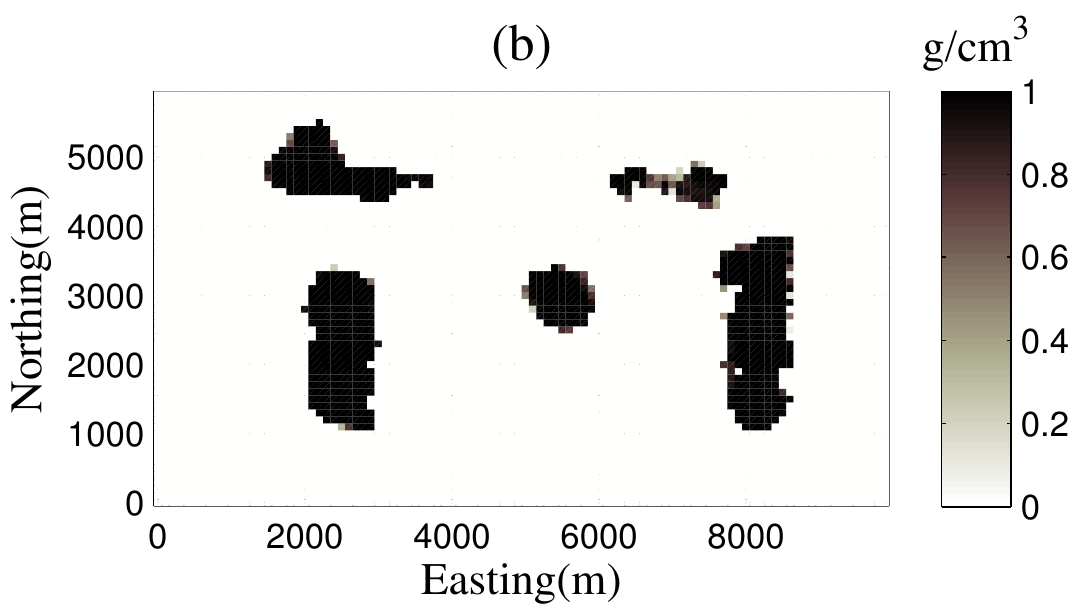}}
\subfigure{\label{18c}\includegraphics[width=.45\textwidth]{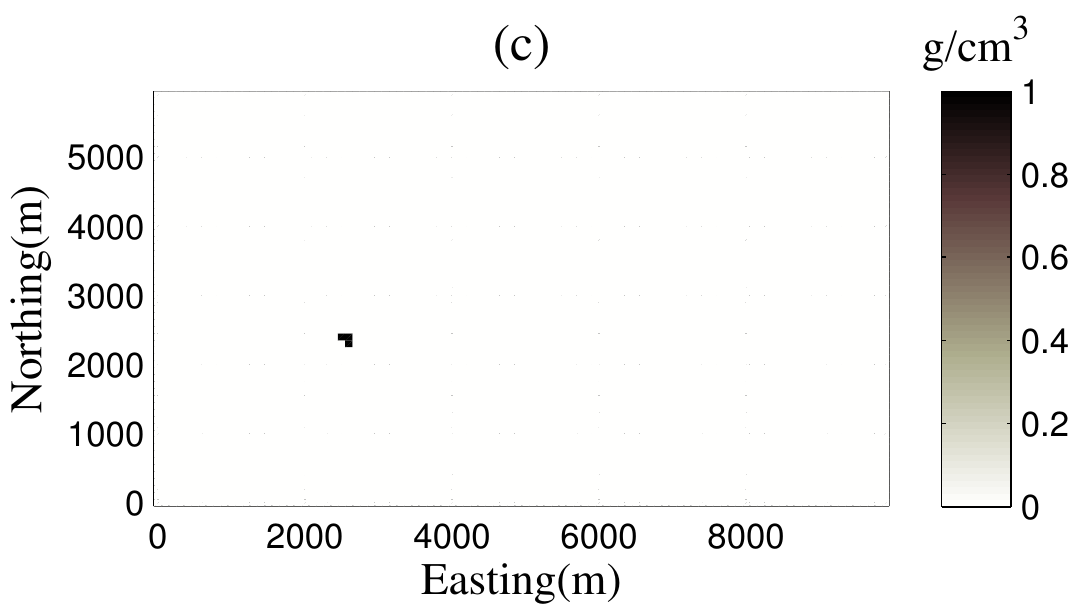}}
\subfigure{\label{18d}\includegraphics[width=.45\textwidth]{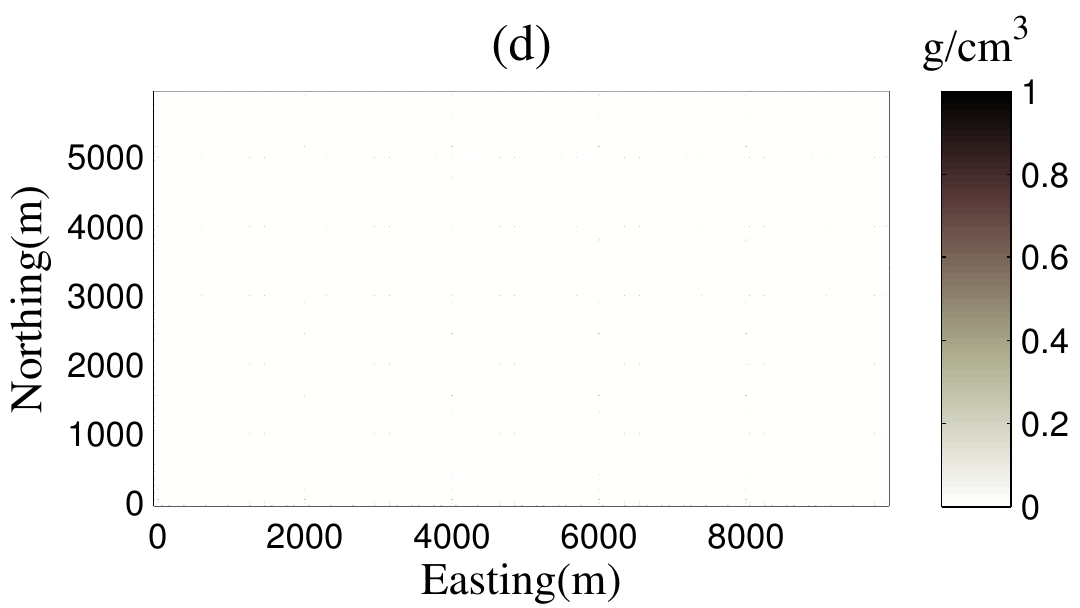}}
\caption{ For the data in Fig.~\ref{13b}: The reconstructed model using Algorithm~\ref{projectedalgorithm} with $t=350$ and  UPRE method. The depths of the sections are: (a) $Z=100$ m; (b) $Z=300$ m; (c) $Z=500$ m and (d)  $Z=700$ m.} \label{fig18}
\end{figure*}

\begin{figure*}
\subfigure{\label{19a}\includegraphics[width=.4\textwidth]{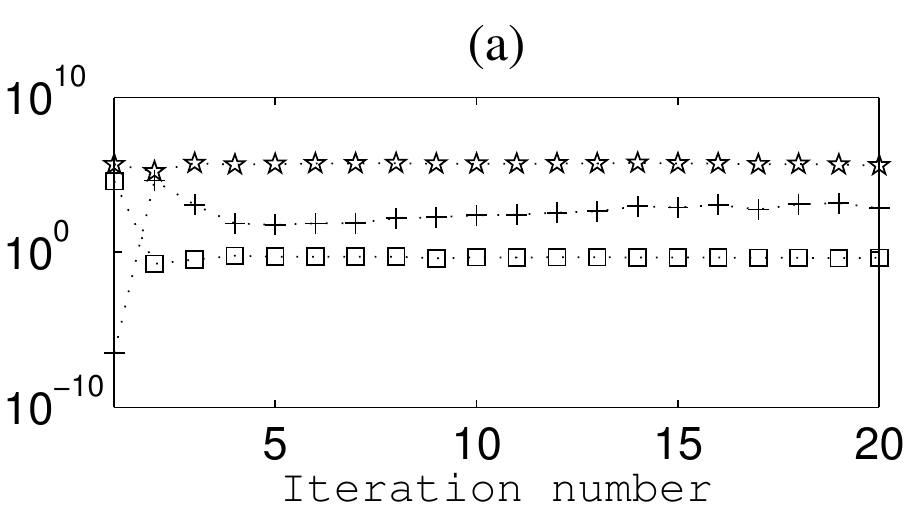}}
\subfigure{\label{19b}\includegraphics[width=.4\textwidth]{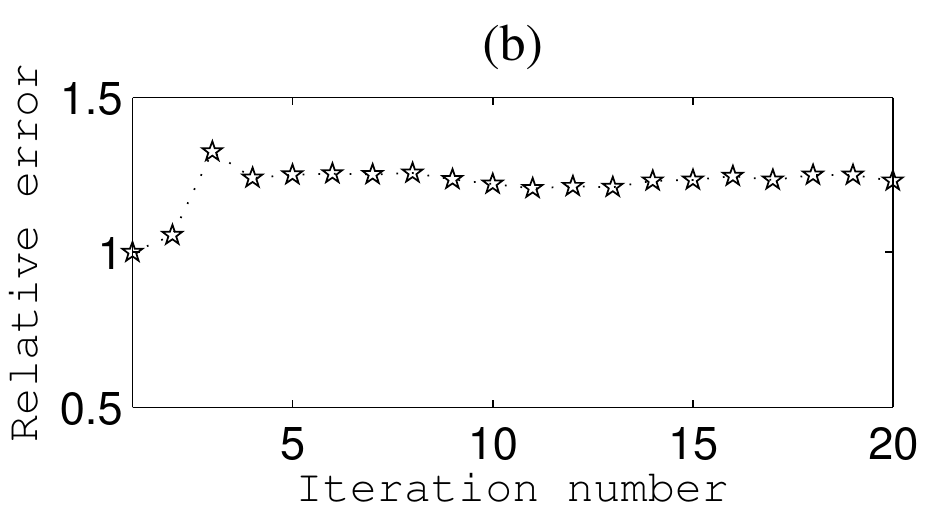}}
\subfigure{\label{19c}\includegraphics[width=.4\textwidth]{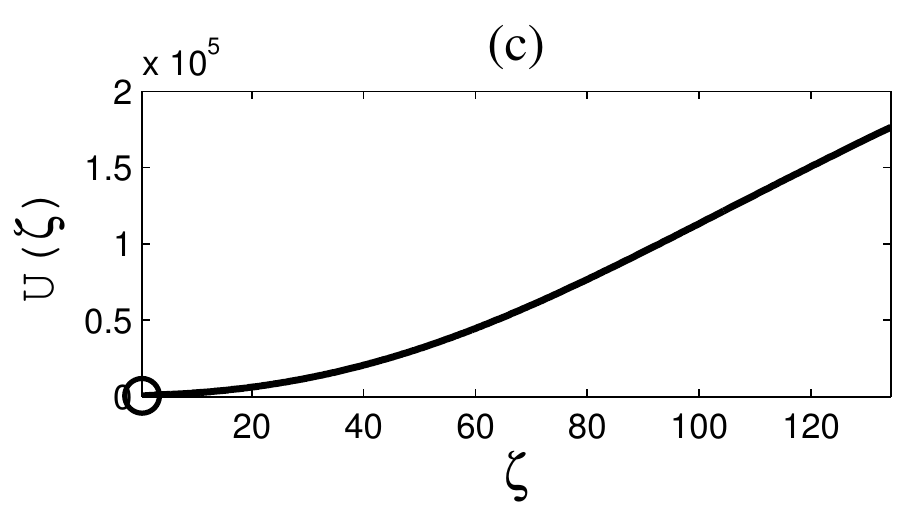}}
\caption {For the reconstructed model in Fig.~\ref{fig18}. In (c) the UPRE function at iteration $20$.} \label{fig19}
\end{figure*}

We now implement Algorithm~\ref{projectedalgorithm} for the projected subspace of size $t=350$  using the UPRE function \refeq[subupre] rather than the TUPRE to find the  regularization parameter. All other parameters are the same as before.  The results are illustrated in Figs.~\ref{fig18} and \ref{fig19}. The recovered model is not reasonable and the relative error is very large. The iterations are terminated at $K_{\mathrm{max}}=20$. This is similar to our results obtained for the cube and demonstrate that truncation of the subspace is required in order to obtain a suitable regularization parameter.   

\section{Real data: Mobrun Anomaly, Noranda, Quebec}\label{real}

\begin{figure*}
\includegraphics[width=0.45\textwidth]{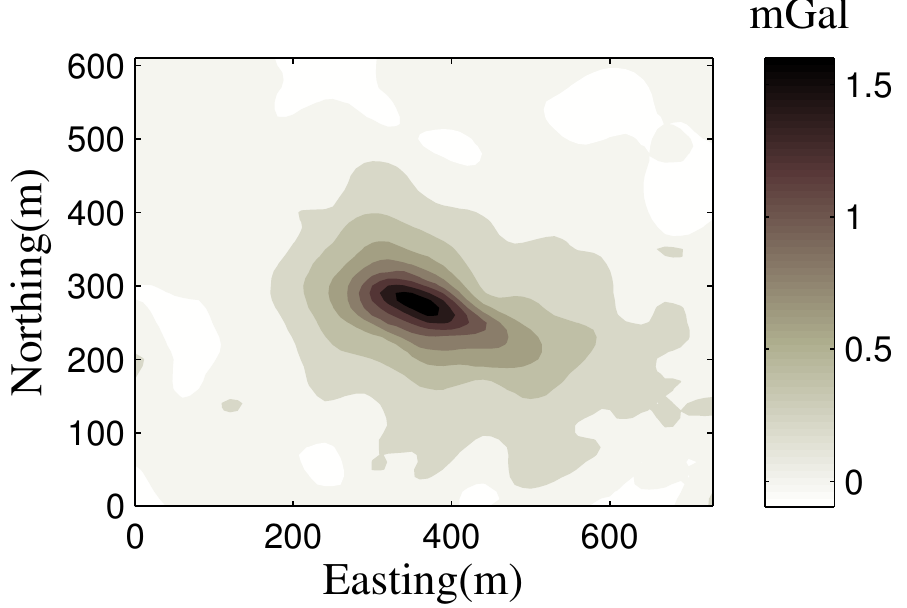}
\caption{Residual Anomaly of Mobrun ore body, Noranda, Quebec, Canada.} \label{fig20}
\end{figure*}

To illustrate the relevance of the approach for a practical case we use the residual gravity data from the well-known Mobrun ore body, northeast of Noranda, Quebec. The anomaly pattern is associated with a massive body of base metal sulfide (mainly pyrite) which has displaced volcanic rocks of middle Precambrian age \cite{GrWe:65}. We carefully digitized the data from Fig. 10.1 in Grant $\&$ West \shortcite{GrWe:65}, and re-gridded onto a regular grid of $74 \times 62 =4588$ data in $x$ and $y$ directions respectively, with grid spacing $10$~m,  see Fig. \ref{fig20}. In this case, the densities of the pyrite and volcanic host rock were taken to be $4.6$~g~cm$^{-3}$ and $2.7$~g~cm$^{-3}$, respectively \cite{GrWe:65}.
For the data inversion we use a model with cells  of width $10$~m in the eastern and northern directions. In the depth dimension the first $10$ layers of cells have a thickness of $5$~m, while the subsequent layers increase  gradually to $10$~m. The maximum depth of the model is $160$~m. This yields a model with the $z$-coordinates: $0:5:50$,  $56$,    $63$,    $71$,    $80:10:160$ and  a mesh  with $ 74 \times 62 \times 22 = 100936 $ cells. We suppose each datum has an error with standard deviation $\left( 0.03 (\bfdo)_i+ 0.004 \|\bfdo\|\right)$. Algorithm~\ref{projectedalgorithm} is used with TUPRE, $t=300$ and $\epsilon^2=1.e{-9}$. 

\begin{figure*}
\subfigure{\label{21a}\includegraphics[width=.45\textwidth]{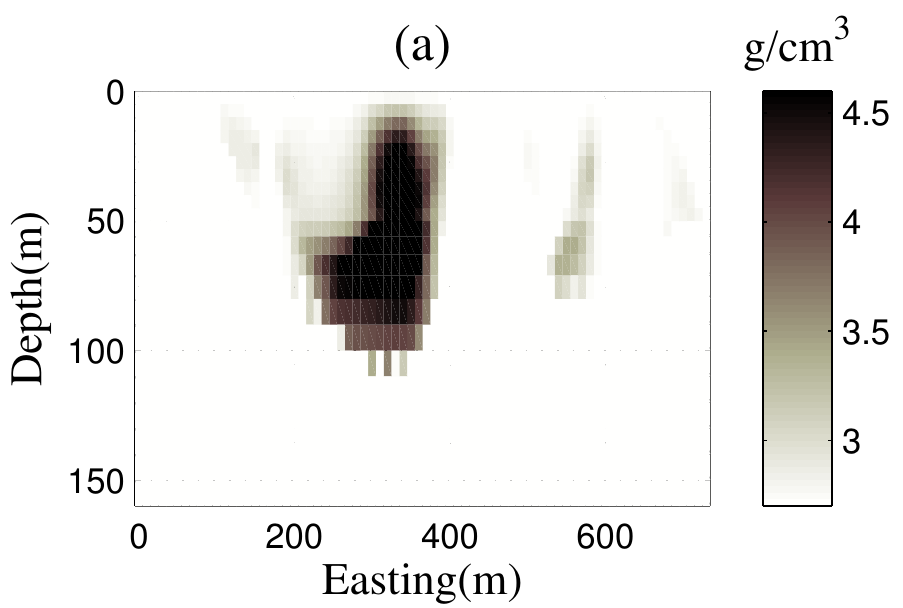}}
\subfigure{\label{21b}\includegraphics[width=.45\textwidth]{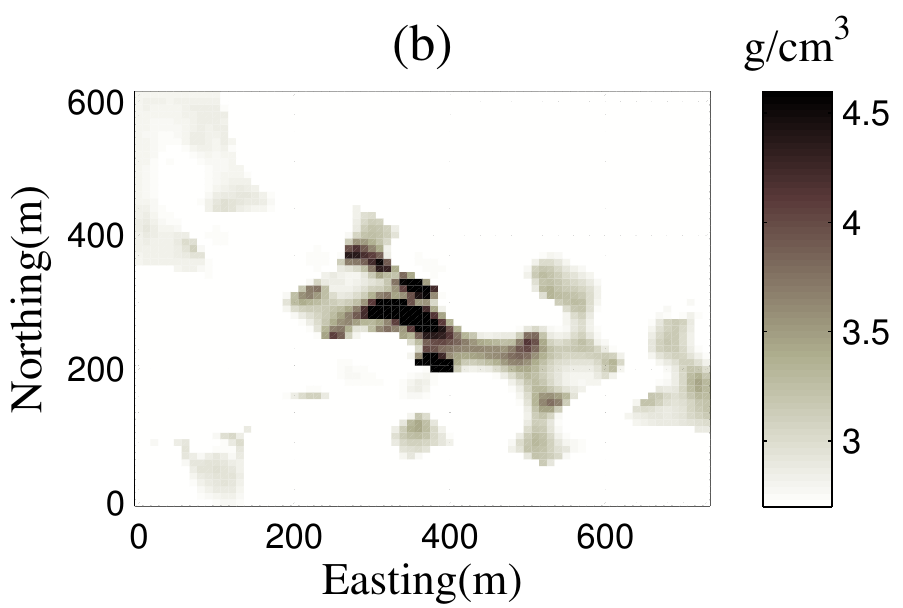}}
\caption {The reconstructed model for the data in Fig.~\ref{fig20} using Algorithm~\ref{projectedalgorithm} with $t=300$ and TUPRE method. (a) The cross-section at $y=285$~m; (b) The  plane-section at $z=50$~m.} \label{fig21}
\end{figure*}

\begin{figure*}
\subfigure{\label{22a}\includegraphics[width=.4\textwidth]{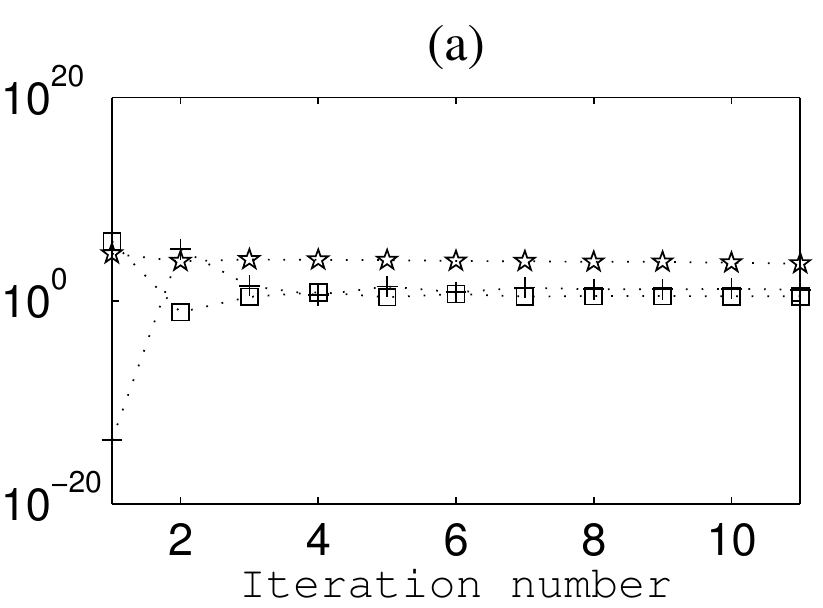}}
\subfigure{\label{22b}\includegraphics[width=.4\textwidth]{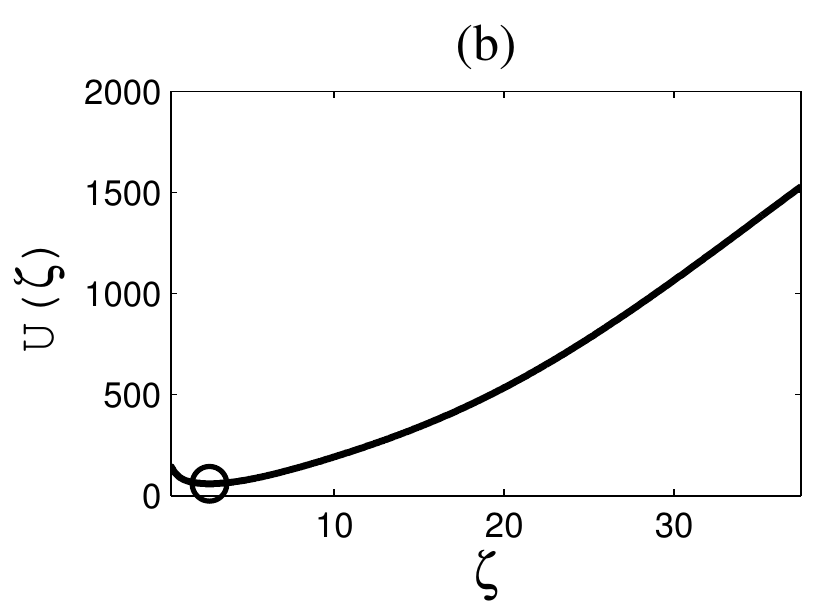}}
\caption {The inversion results for the reconstructed model in Fig.~\ref{fig21}. In (b) the TUPRE function at iteration $11$.} \label{fig22}
\end{figure*}

The inversion process terminates after $11$ iterations. The cross-section of the recovered model at $y=285$~m and a plane-section at $z=50$~m are shown in Figs.~\ref{21a} and ~\ref{21b}, respectively. The depth to the surface is about $10$ to $15$~m, and the body extends to a maximum depth of $110$~m. The results are in good agreement with those obtained from previous investigations and drill hole information, especially for depth to the surface and intermediate depth, see Figs. $10$-$23$ and $10$-$24$ in Grant $\&$ West \shortcite{GrWe:65}, in which they interpreted the body with about $305$~m in length, slightly more than $30$~m in maximum width and having a maximum depth of $183$~m. It was confirmed by the drilling that no dense material there is at deeper depth \cite{GrWe:65}. To compare with the results of other inversion algorithms in this area, we suggest the paper by Ialongo et al. \shortcite{IaFeFl:14}, where they illustrated the inversion results for the gravity and first-order vertical derivative of the gravity (\cite[Figs. $14$ and $16$]{IaFeFl:14}). The algorithm is fast, requiring less than $30$ minutes, and yields a model with blocky features. The progression of the data misfit, the regularization term and the regularization parameter with iteration $k$ and the TUPRE function at the final iteration  are shown in Figs.~\ref{22a} and ~\ref{22b}.
 
\section{Conclusions}\label{conclusion}
An algorithm for inversion of gravity data using iterative $L_1$ stabilization has been presented. The linear least squares problem at each iteration is solved on the projected space generated using Golub-Kahan bidiagonalization. Using the  UPRE method to estimate  the regularization parameter for the subspace solution underestimates the regularization parameter and thus provides solutions that are not satisfactory. This occurs because the sensitivity matrix  of the gravity inversion problem is only mildly, or at worst, moderately ill-conditioned. Thus, the singular values of the projected matrix can not all approximate large singular values of the original matrix exactly. Instead the spectrum of the projected problem inherits the ill-conditioning of the full problem and only accurately approximates a portion of the dominant spectrum of the full problem. This leads to underestimating the regularization parameter. We used different noise levels for the synthetic data and showed that the problem is worse for higher noise levels. We demonstrated that using a truncated projected spectrum gives an effective regularization parameter. The new method, here denoted as TUPRE, gives results using the projected subspace algorithm which are comparable with those obtained for the full space, while just requiring the generation of a small projected space. The presented algorithm is practical and efficient and has been illustrated for the inversion of  synthetic and real gravity data. Our results showed that the gravity inverse problem can be efficiently and effectively inverted using the GKB projection with regularization applied on the projected space. By numerical examination, we have suggested how to determine the size of the projected space and the truncation level based on the number of measurements $m$.  This provides a simple, but practical, rule which can be used confidently. Furthermore, while the examples used in the paper are from small to moderate size, and the code is implemented on a desktop computer, we believe for very large problems the situation is the same and truncation is required for parameter estimation in conjunction with using the LSQR algorithm.

\begin{acknowledgments}
Rosemary Renaut acknowledges the support of  NSF grant  DMS 1418377:   ``Novel Regularization for Joint Inversion of Nonlinear Problems". 
\end{acknowledgments}

\appendix
\section{solution using singular value decomposition}\label{svdsolution}
Suppose  $m^* = min (m,n)$ and the SVD of matrix $ \tilde{\tilde{G}} \in \Rmn $ is given by $\tilde{\tilde{G}}=U \Sigma V^T$, where the singular values are ordered $\sigma_1 \geq \sigma_2\geq \dots \geq \sigma_{m^*} > 0$,  and occur
on the diagonal of $\Sigma \in \Rmn$ with $n-m$ zero columns (when $m <n $) or $m-n$ zero rows (when $m>n$), using the full definition of the SVD, \cite{GoLo:96}. $U \in \mathcal{R}^{m \times m} $, and $ V \in \mathcal{R}^{n \times n}$ are orthogonal matrices with columns denoted by  $\bfui$ and $\bfvi$. Then the solution of (\ref{globalfunctionh}) is given by
\begin{eqnarray}\label{hsvd}
\bfh(\alpha)= \sum_{i=1}^{m^*} \frac{\sigma_i^2}{\sigma_i^2 +\alpha^2} \frac{\bfui^T \tilde{\bfr}}{\sigma_i} \bfvi.
\end{eqnarray}
For the projected problem $B_t \in \mathcal{R} ^{(t+1) \times t}$, i.e. $m >n$, and the expression still applies to give  the solution of (\ref{zsolution}) with $\| \tilde{\bfr} \|_2 \bfe$ replacing $\tilde{\bfr}$,  $\zeta$ replacing $\alpha$,  $\gamma_i$ replacing $\sigma_i$ and $m^* =t$ in (\ref{hsvd}).

\section{UPRE function using SVD}\label{svdparameter}
The UPRE function for determining $\alpha$ in the Tikhonov form \refeq[globalfunctionh] with system matrix  $ \tilde{\tilde{G}}$ is expressible using the SVD for $\tilde{\tilde{G}}$
\begin{eqnarray*}
U(\alpha)=\sum_{i=1}^{m^*}  \left( \frac{1}{\sigma_i^2 \alpha^{-2} + 1} \right)^2 \left(\bfui^T\tilde{\bfr} \right)^2 + 2 \left( \sum_{i=1}^{m^*} \frac{\sigma_i^2}{\sigma_i^2+\alpha^2}\right) - m.
\end{eqnarray*}
In the same way the UPRE function for the projected problem \refeq[globalfunctionproj] is given by
\begin{eqnarray*}
U(\zeta)=\sum_{i=1}^{t}  \left( \frac{1}{\gamma_i^2 \zeta^{-2} + 1} \right)^2 \left(\bfui^T(\| \tilde{\bfr} \|_2 \bfe) \right)^2 + \sum_{i=t+1}^{t+1}\left( \bfui^T(\| \tilde{\bfr} \|_2 \bfe) \right) ^2 +2 \left( \sum_{i=1}^{t} \frac{\gamma_i^2}{\gamma_i^2+\zeta^2}\right) - (t+1). 
\end{eqnarray*}
Then, for truncated UPRE, $t$  is  replaced by $t_{\mathrm{trunc}} < t$ so that the terms from $t_{\mathrm{trunc}}$ to $t$ are ignored, corresponding to dealing with these as constant with respect to the minimization of $U(\zeta)$.

\section{Golub-Kahan Bidiagonalization}\label{gkbappen}
The Golub-Kahan Bidiagonalization (GKB) algorithm starts with the right-hand side $\tilde{\bfr}$ and matrix $\tilde{\tilde{G}}$, takes the following simple steps, see \cite{Hansen:2007,Hansen:2010} , in which the quantities $\alpha_{t}$ and $\beta_{t}$ are chosen such that the corresponding vectors $a_{t}$ and $h_{t}$ are normalized: 
\begin{algorithm}
\begin{algorithmic}[]
\STATE $a_{0}=0$, $\beta_{1}=||\tilde{\bfr}||_2 $, $h_{1}=\tilde{\bfr}/\beta_{1}$
\STATE $\mathrm{for}$ {$ t=1,2,... $} 
\STATE $\alpha_{t} a_{t}=\tilde{\tilde{G}}^T h_{t} -\beta_{t} a_{t-1} $
\STATE $\beta_{t+1} h_{t+1} = \tilde{\tilde{G}} a_{t} - \alpha_{t} h_{t} $
\STATE $\mathrm{end}$
\end{algorithmic}
\end{algorithm}

After $t$ iterations, this simple algorithm produces two matrices $A_t \in \mathcal{R}^{n \times t}$ and $H_{t+1} \in \mathcal{R}^{m \times (t+1)}$ with orthonormal columns,
\begin{eqnarray*}
A_t=(a_1,a_2,...,a_t), \quad H_{t+1}=(h_1,h_2,...,h_{t+1})
\end{eqnarray*}
and a lower bidiagonal matrix $ B_t \in \mathcal{R}^{(t+1) \times t}$ ,
\begin{eqnarray*}
B_t=
  \begin{bmatrix}
    \alpha_1&          &  &  \\
    \beta_2 & \alpha_2 &  &  \\
            & \beta_3 & \ddots  &  \\
            &         &  \ddots & \alpha_t  \\
            &         &         & \beta_{t+1} \\
  \end{bmatrix}
\end{eqnarray*}
such that 
\begin{eqnarray}
\tilde{\tilde{G}}A_t = H_{t+1}B_t, \quad  H_{t+1}\bfe=\tilde{\bfr}/\| \tilde{\bfr}\|_2.
\end{eqnarray}
Where $\bfe$ is the unit vector of length $t+1$ with a $1$ in the first entry. The columns of $A_t$ are the desired basis vectors for the Krylov subspace $\mathcal{K}_t$, (\ref{krylov}).

\end{document}